\documentclass[twocolumn, tighten]{aastex631}
\usepackage{url}
\usepackage{amsmath}
\usepackage{multirow}
\usepackage{booktabs}
\usepackage{hyperref}
\usepackage{latexsym}
\received{--}
\revised{--}
\accepted{--}
\submitjournal{ApJ}
\shorttitle{SN~2017iro}
\shortauthors{B. Kumar et al.}
\begin{document}

\title{Investigating the observational properties of Type Ib supernova SN~2017iro}

\correspondingauthor{Brajesh Kumar}
\email{brajesh@aries.res.in, brajesharies@gmail.com}

\author[0000-0001-7225-2475]{Brajesh Kumar}
\affiliation{Aryabhatta Research Institute of Observational Sciences, Manora Peak, Nainital - 263001, India}
\affiliation{Indian Institute of Astrophysics, II Block, Koramangala, Bengaluru 560034, India}

\author[0000-0003-2091-622X]{Avinash Singh}
\affiliation{Hiroshima Astrophysical Science Center, Hiroshima University, Higashi-Hiroshima, Hiroshima 739-8526, Japan}
\affiliation{Indian Institute of Astrophysics, II Block, Koramangala, Bengaluru 560034, India}

\author[0000-0002-6688-0800]{D.K. Sahu}
\affiliation{Indian Institute of Astrophysics, II Block, Koramangala, Bengaluru 560034, India}

\author[0000-0003-3533-7183]{G.C. Anupama}
\affiliation{Indian Institute of Astrophysics, II Block, Koramangala, Bengaluru 560034, India}

\begin{abstract}

We report results of optical imaging and low-resolution spectroscopic monitoring of supernova (SN)~2017iro that occurred in the nearby ($\sim$\,31 Mpc) galaxy NGC~5480. The \ion{He}{1} 5876 \AA\, feature present in the earliest spectrum (--\,7 d) classified it as a Type Ib SN. The follow-up observations span from --\,7 to +\,266 d with respect to the $B$-band maximum. With a peak absolute magnitude in $V$-band, ($M_{V}$)\,=\,$-17.76\pm0.15$ mag and bolometric luminosity (log$_{10}$\,L)\,=\,42.39\,$\pm$\, 0.09 erg s$^{-1}$, SN~2017iro is a moderately luminous Type Ib SN. The overall light curve evolution of SN~2017iro is similar to SN~2012au and SN~2009jf during the early (up to $\sim$100 d) and late phases ($>$150 d), respectively. The line velocities of both \ion{Fe}{2} 5169 \AA\, and \ion{He}{1} 5876 \AA\, are $\sim$\,9000 km s$^{-1}$ near the peak. The analysis of the nebular phase spectrum ($\sim$\,+209 d) indicates an oxygen mass of $\sim$\,0.35 M$_{\odot}$. The smaller [\ion{O}{1}]/[\ion{Ca}{2}] flux ratio of $\sim$\,1 favours a progenitor with a zero-age main-sequence mass in the range $\sim$\,13--15 M$_{\odot}$, most likely in a binary system, similar to the case of iPTF13bvn. The explosion parameters are estimated by applying different analytical models to the quasi-bolometric light curve of SN~2017iro. $^{56}$Ni mass synthesized in the explosion has a range of $\sim$\,0.05\,--\,0.10 M$_{\odot}$, the ejecta mass $\sim$1.4\,--\,4.3 M$_{\odot}$ and the kinetic energy $\sim$\,0.8\,--\,1.9\,$\times$ 10$^{51}$ erg.

\end{abstract}


\section{Introduction}\label{intro}

Type Ib, Ic, Ic-broad lined (Ic-BL), and IIb supernovae (SNe) belong to the class of stripped-envelope supernovae \citep[SE-SNe,][]{1941PASP...53..224M, 1997ApJ...491..375C, 1997ARA&A..35..309F, 2017Galyam, 2019Modjaz}. SE-SNe are generally characterized by bell-shaped light curves (LCs) which are mainly powered by the thermal energy released as a consequence of radioactive decay of the iron group elements ($^{56}$Ni $\rightarrow$ $^{56}$Co $\rightarrow$ $^{56}$Fe) synthesized in the explosion. SE-SNe show no hydrogen (Ib) or neither hydrogen nor helium (Ic) in their spectra except for Type IIb events, which exhibit hydrogen in the early phase spectra \citep*{1988AJ.....96.1941F, 1993Filippenko}. This indicates that they are the results of explosion of stars that are devoid of the outer layer of hydrogen (Ib) or both hydrogen and helium (Ic) at the time of explosion. The spectroscopic behaviour of SE-SNe separates them from Type II SNe, where hydrogen is found to be dominant in the spectra. The SE-SNe and Type II events are collectively grouped as core-collapse supernovae (CC-SNe) as they result from the gravitational collapse of the iron core of massive stars ($M \ge$ 8\, M$_{\sun}$) during the end stages of their lives \citep{2003ApJ...591..288H,2009ARA&A..47...63S}.

In SE-SNe, the mechanisms responsible for removing the outer envelopes (hydrogen or helium) of the progenitors can be either radiation-driven stellar winds \citep*{1975MSRSL...9..193C, 2008A&ARv..16..209P, 2012pauldrach}, eruptive mass loss \citep{2006ApJ...645L..45S, 2006smith} and/or mass transfer to a companion star \citep*{1984nomoto, 1987nomoto, 1985ApJ...294L..17W, 1992ApJ...391..246P,1995woosley, 1995PhR...256..173N, 1999wellstein, Wellstein-2001, 2002pols, 2004pods, 2010ApJ...725..940Y}. Depending on the main-sequence mass of the progenitor star, above mechanisms or a combination of them operates. In the case of a single massive ($>$\,20 M$_{\sun}$) Wolf-Rayet star, the radiation-driven stellar wind is the dominant process \citep[see][and references therein]{Woosley-1995, Massey-2003, Crowther-2007, Smith-2014, Yoon-2015AA}. In case of moderately massive (8\,--\,25 M$_{\sun}$) stars, mass transfer to the binary companion effectively removes the outer envelope \citep*[][and references therein]{2013A&A...558L...1G, Groh-2013, 2014ARA&A..52..487S, 2015PASA...32...15Y}. 

Dedicated and large area transient survey programmes have made it possible to discover a large number of SNe of different types. The SE-SNe constitute a reasonable fraction ($\sim$25\,--\,30\%) of the overall SN rate in the local Universe \citep{2011MNRAS.412.1441L,2013MNRAS.436..774E,2017Graur-1,2017Graur-2}. The availability of ample data has led to a qualitative analysis of a large sample of SE-SNe in the recent years \citep[see, e.g.][]{2011ApJ...741...97D, 2014ApJS..213...19B, 2014AJ....147...99M,2015A&A...574A..60T,2016ApJ...827...90L,2016MNRAS.457..328L,2016MNRAS.458.2973P,2018AA...609A.136T,2019MNRAS.485.1559P, 2019Shivvers, 2019williamson, 2021shahbandeh}. These studies reveal that heterogeneity exists in the observational properties of SE-SNe in terms of synthesized $^{56}$Ni masses ($\sim$0.1\,--\,0.4 M$_{\sun}$), ejecta masses ($\sim$1\,--\,6 M$_{\sun}$), explosion energies (few 10$^{51}$ erg), and absolute magnitude ($\sim$--17 to --18 mag). Notably, the sample selection criteria and the methods applied may induce inconsistency in various estimates \citep[see,][]{2018AA...609A.136T,2019MNRAS.485.1559P}.

Statistical investigations of a large sample of photometric data are useful in exploring the bulk properties of various types of events. Nevertheless, such studies are generally limited to LCs in selected pass-bands, and also, the follow-up covers a short duration of the SN evolution. Further, the spectroscopic follow-up of most objects is restricted to early phases. Non-uniformity in the data sample may also be present as these are collected at different observing facilities and detectors. Studying individual events with proper monitoring at different phases (both photospheric and nebular) is hence extremely important. The very early phase observations (hours to days after explosion) of these events are useful to constrain the progenitor radius at its end stage. This needs a very early detection and quick follow up which is not always possible considering their random occurrence in the sky. The large area surveys with high cadence have contributed significantly in this regard. SN~1993J (Type IIb) is the first event among SE-SNe which shows evidence of a prominent cooling tail just after explosion \citep{1994-Richmond,1995-Barbon}. During last two decades several Type IIb SNe with such interesting features have been monitored and studied well e.g. SN~2008ax \citep{2008-Pastorello,2009-Roming}, SN~2011dh \citep{2011-Arcavi}, SN~2011fu \citep{2013-Kumar,2015-Garoffolo}, SN~2013df \citep{2014A&A...565A.114F} and SN~2016gkg \citep{2017-Arcavi,2017-Tartaglia}. A handful of Type Ib events have also been discovered at very early phases and their observational properties are studied in detail such as SN~1999ex \citep{2002-Hamuy, 2002AJ....124.2100S}, SN~2008D \citep{2008-Soderberg, 2008-Mazzali, 2009-Malesani, 2009-Modjaz,2013-Bersten}, iPTF13bvn \citep{2016-Fremling, 2014-Bersten} and LSQ13abf \citep{2020-Stritzinger}. Recently, \citet{2019MNRAS.485.1559P} analyzed the properties of 18 SE-SNe and discussed the implications for their progenitors. Direct detection of SE-SNe progenitor candidates has been possible only for a few objects, e.g. SN~1993J \citep{2004Natur.427..129M}, SN~2011dh \citep{2011ApJ...739L..37M,2013ApJ...772L..32V}, SN~2001ig \citep{2018-Ryder}, iPTF13bvn \citep{2013ApJ...775L...7C, 2015-Eldridge, 2015kuncarayakti, 2016ApJ...825L..22F}, SN~2016gkg \citep{2017-Kilpatrick, 2017-Tartaglia}, SN~2017ein \citep{2018-Kilpatrick, 2018-VanDyk, 2019-Xiang} and SN~2019yvr \citep{2021-Kilpatrick}. Along with very early phases, the temporal observations during maximum to nebular phases are equally important to estimate various explosion parameters and progenitor properties. Detail investigation of more events can provide an alternative way to understand various progenitor channels. 

SN~2017iro was discovered by Patrick Wiggins on 2017 November 30.5 UT \citep{2017TNSTR1354....1W} at about 15 arcsec South-East of the centre of the galaxy NGC~5480 which also hosted another Type Ib event SN~1988L \citep{1988IAUC.4590....1P}. Pre-explosion images of host galaxy, obtained by \textit{Hubble Space Telescope} (HST) in F606W-band using Wide Field Camera (WFC) of Advanced Camera for Surveys (ACS), exist in the HST archive. SN~2017iro was discovered at a magnitude of $\sim$15.5 mag (clear filter) and classified as a Type Ib/c SN with the help of the spectrum taken on 2017 December 02 \citep{2017Bertrand}. To be noted that SN~2017iro discovery was slightly late ($\sim$\,8 d after explosion) however, we were able to cover its peak as well as the late phase observations with a good sampling of photometric and spectroscopic data points. Table~\ref{tab_1} lists details of SN~2017iro and its host galaxy.

Results based on well-sampled optical photometric and low-resolution spectroscopic observations of SN~2017iro are presented in this paper. The observations and data reduction are described in Section~\ref{obs}. Light curve and spectral properties are presented in Section~\ref{lc} and \ref{spec_anal}, respectively. In Section~\ref{diss}, the results are discussed, and the summary is provided in Section~\ref{sum}.

\begin{figure}
\centering
\includegraphics[width=\columnwidth]{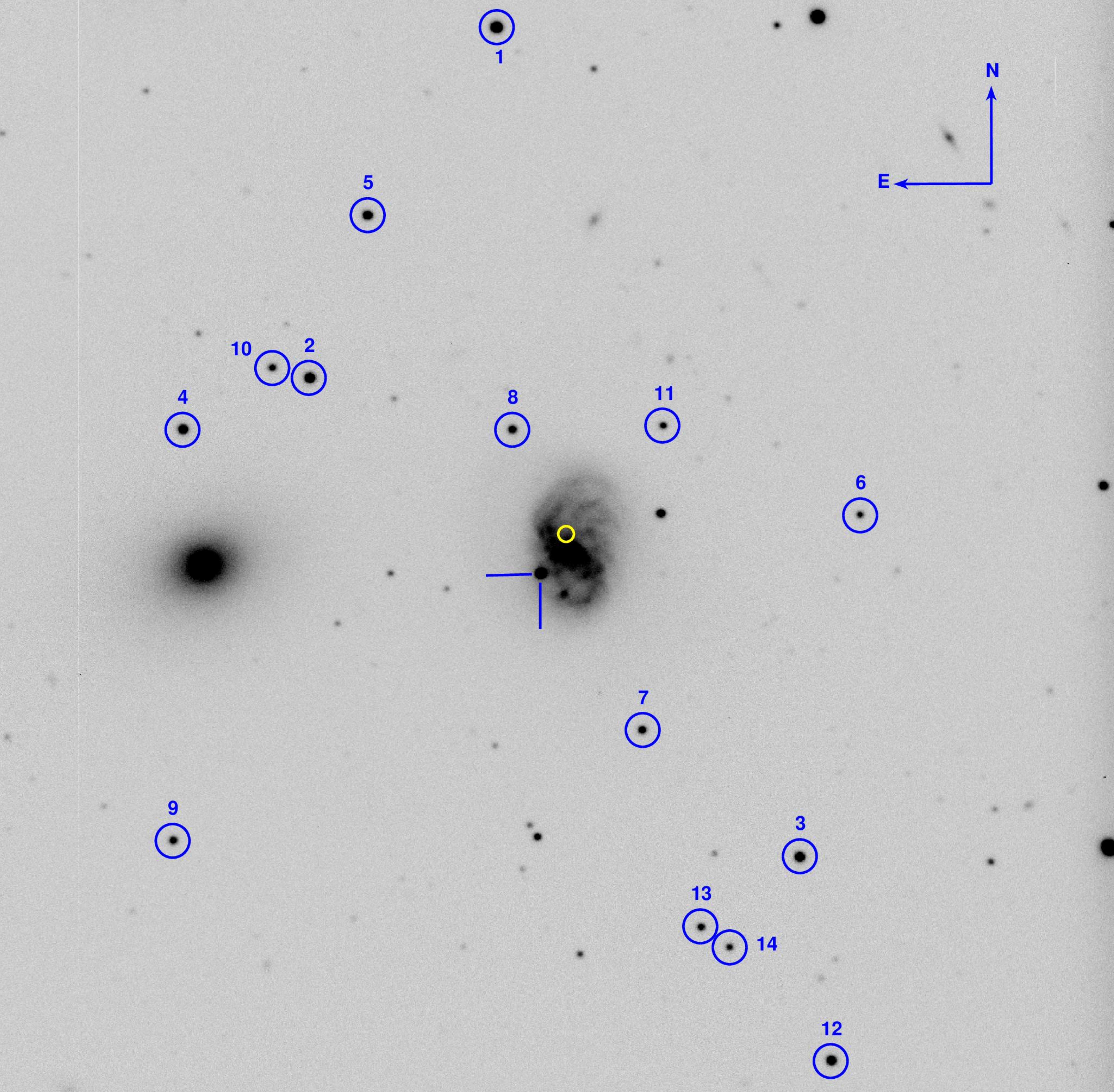}
\caption{Identification chart of SN~2017iro obtained in $V$-band (SN is marked with blue crosshairs). The field of view is roughly 9 arcmin $\times$ 9 arcmin. The IDs 1--14 indicate secondary standard stars in the SN field (their magnitudes are listed in Table~\ref{tab_id}). The location of SN~1988L is also marked with a bold circle (yellow).} 
\label{fig_1}
\end{figure}

\begin{table}
\scriptsize
\centering
\caption{Details of SN~2017iro and its host galaxy.}\label{tab_1}
\begin{tabular}{llc}
\hline
Parameters                         &   Value                            &  Reference        \\ 
\hline
SN~2017iro:                        &                                    &                   \\    
\quad RA (J2000)                   &   14$^{\rm h}$~06$^{\rm m}$~23\fs11& 1                 \\   
\quad DEC (J2000)                  &  +50\degr ~43\arcmin ~20\farcs20   & 1                 \\ 
\quad Explosion epoch {\rm (UT)}   &   2017 November 23                 & Section~\ref{expl}\\
                                   &   (JD 2458080.7)                   &                   \\
\quad Discovery date (UT)          &   2017 November 30.5               & 1                 \\
                                   &   (JD 2458088.0)                   &                   \\
\quad $E(B-V)_{total}$             &   0.28 $\pm$ 0.04 mag        & Section~\ref{color}     \\\\ 
NGC~5480:                          &                                    &                   \\
\quad Type                         &   SA(s)c                           & 2                 \\
\quad Redshift                     &   0.006191 $\pm$ 0.000017          & 2                 \\
\quad Distance                     &   30.8 $\pm$ 2.2 Mpc               & 3                 \\
\quad Distance modulus             &   32.44 $\pm$ 0.15 mag             & 3                 \\ 
\hline
\end{tabular}  
\begin{minipage}{\textwidth}
$^{1}$ \url{https://wis-tns.weizmann.ac.il/object/2017iro} \\
$^{2}$ NASA/IPAC Extragalactic Database (NED) \\
$^{3}$ \citet{2000-Mould}
\end{minipage}
\end{table}
\raggedbottom


\begin{table*}
\scriptsize
\centering
\caption{Calibrated magnitudes of secondary standards in the field of SN~2017iro.
Star IDs are indicated in Fig.~\ref{fig_1}.}\label{tab_id}
\begin{tabular}{cccccc}
\hline
Star & $U$              & $B$              & $V$              & $R$              & $I$              \\
ID   & (mag)            & (mag)            & (mag)            & (mag)            & (mag)            \\ \hline
1    & 15.86 $\pm$ 0.07 & 15.93 $\pm$ 0.02 & 15.34 $\pm$ 0.01 & 14.95 $\pm$ 0.02 & 14.56 $\pm$ 0.01 \\
2    & 16.16 $\pm$ 0.06 & 16.39 $\pm$ 0.01 & 15.87 $\pm$ 0.01 & 15.53 $\pm$ 0.02 & 15.20 $\pm$ 0.01 \\
3    & 18.02 $\pm$ 0.02 & 17.21 $\pm$ 0.02 & 16.21 $\pm$ 0.01 & 15.59 $\pm$ 0.01 & 15.07 $\pm$ 0.01 \\
4    & 18.09 $\pm$ 0.09 & 17.26 $\pm$ 0.01 & 16.22 $\pm$ 0.01 & 15.57 $\pm$ 0.02 & 15.00 $\pm$ 0.01 \\
5    & 17.21 $\pm$ 0.06 & 17.10 $\pm$ 0.01 & 16.41 $\pm$ 0.01 & 15.97 $\pm$ 0.02 & 15.54 $\pm$ 0.01 \\
6    & 19.83 $\pm$ 0.05 & 19.33 $\pm$ 0.05 & 17.85 $\pm$ 0.02 & 16.50 $\pm$ 0.02 & 14.91 $\pm$ 0.02 \\
7    & 17.62 $\pm$ 0.01 & 17.74 $\pm$ 0.01 & 17.14 $\pm$ 0.01 & 16.76 $\pm$ 0.02 & 16.41 $\pm$ 0.02 \\
8    & 17.33 $\pm$ 0.04 & 17.51 $\pm$ 0.02 & 16.95 $\pm$ 0.02 & 16.59 $\pm$ 0.03 & 16.24 $\pm$ 0.03 \\
9    & 18.14 $\pm$ 0.02 & 17.96 $\pm$ 0.01 & 17.23 $\pm$ 0.01 & 16.78 $\pm$ 0.02 & 16.38 $\pm$ 0.01 \\
10   & 19.92 $\pm$ 0.69 & 18.90 $\pm$ 0.04 & 17.48 $\pm$ 0.02 & 16.30 $\pm$ 0.02 & 14.92 $\pm$ 0.01 \\
11   & 18.27 $\pm$ 0.01 & 18.33 $\pm$ 0.02 & 17.68 $\pm$ 0.02 & 17.28 $\pm$ 0.02 & 16.84 $\pm$ 0.06 \\
12   & 18.17 $\pm$ 0.05 & 17.36 $\pm$ 0.01 & 16.40 $\pm$ 0.01 & 15.80 $\pm$ 0.02 & 15.29 $\pm$ 0.01 \\
13   & 18.04 $\pm$ 0.01 & 18.11 $\pm$ 0.01 & 17.51 $\pm$ 0.01 & 17.12 $\pm$ 0.02 & 16.74 $\pm$ 0.01 \\
14   & 18.66 $\pm$ 0.02 & 18.52 $\pm$ 0.02 & 17.80 $\pm$ 0.02 & 17.36 $\pm$ 0.03 & 16.95 $\pm$ 0.03 \\
\hline
\end{tabular}
\end{table*}

\section{Observations and data reduction}\label{obs}
\subsection{Photometric observation}\label{obs_ph}

The photometric follow-up of SN~2017iro started just after the announcement of its discovery, i.e., 2017 December 01 (JD~2458089.5), and continued up to 2018 August 31 (JD~2458362.1). These observations were performed in Bessell $UBVRI$ bands with the Himalayan Faint Object Spectrograph Camera (HFOSC), mounted on the $f/9$ Cassegrain focus of the 2 m Himalayan Chandra Telescope (HCT) of Indian Astronomical Observatory (IAO), Hanle, India \citep{2014Prabhu}. HFOSC is equipped with a liquid nitrogen-cooled 2k $\times$ 4k pixels SITe CCD chip (pixel size 15 $\times$ 15 $\mu$m). The gain and readout noise of the detector are 1.22 e$^{-}$/ADU and 4.87 e$^{-}$, respectively. With a plate scale of 0.296 arcsec per pixel, the central 2k $\times$ 2k region covers a field of 10$\arcmin$\,$\times$\,10$\arcmin$ on the sky and is used for imaging. Further description on the HCT and the HFOSC instrument can be obtained from \url{https://www.iiap.res.in/?q=iao_about}.

During the observations, several bias and twilight flat frames were obtained along with science frames. The usual pre-processing steps, such as bias-subtraction, flat-fielding correction, and cosmic ray removal, were applied to raw images of the SN. For this purpose, standard tasks available in the data reduction software {\small IRAF}\footnote{{\small IRAF} stands for Image Reduction and Analysis Facility. It is distributed by the National Optical Astronomy Observatory, which is operated by the Association of Universities for Research in Astronomy (AURA) under a cooperative agreement with the National Science Foundation.} were used. In order to achieve a better signal-to-noise ratio, multiple frames were taken on some nights and co-added in respective bands after the alignment of the images.

\begin{table*}
\scriptsize
\centering
\caption{Log of photometric observations and estimated $UBVRI$ magnitudes of SN~2017iro.}
\label{tab_hct}
\begin{tabular}{cccccccc} 
\hline
Date        &  JD      & Phase$^{\ast}$ &  $U$               & $B$                & $V$                & $R$                & $I$    \\     
(yyyy-mm-dd)& 2450000+ &(d)             &  (mag)             & (mag)              & (mag)              & (mag)              & (mag)  \\  
\hline   
2017-12-01  & 8089.5   & --\,6.7   &  15.85 $\pm$ 0.11  & 16.29 $\pm$ 0.03 & 16.01 $\pm$ 0.11 & 15.78 $\pm$ 0.04 & 15.67 $\pm$ 0.07  \\  
2017-12-02  & 8090.5   & --\,5.7   &  15.71 $\pm$ 0.11  & 16.20 $\pm$ 0.13 & 15.93 $\pm$ 0.11 & 15.71 $\pm$ 0.03 & 15.56 $\pm$ 0.08  \\  
2017-12-05  & 8093.4   & --\,2.8   &  15.64 $\pm$ 0.11  & 15.96 $\pm$ 0.13 & 15.68 $\pm$ 0.05 & 15.47 $\pm$ 0.09 & 15.33 $\pm$ 0.11  \\
2017-12-06  & 8094.4   & --\,1.8   &  15.64 $\pm$ 0.11  & 16.01 $\pm$ 0.09 & 15.63 $\pm$ 0.11 & 15.38 $\pm$ 0.05 & 15.21 $\pm$ 0.07  \\  
2017-12-07  & 8095.5   & --\,0.7   &  15.63 $\pm$ 0.12  & 15.95 $\pm$ 0.05 & 15.56 $\pm$ 0.03 & 15.25 $\pm$ 0.02 & 15.118 $\pm$ 0.03 \\  
2017-12-08  & 8096.5   & +\,0.3    &  15.68 $\pm$ 0.12  & 15.92 $\pm$ 0.06 & 15.56 $\pm$ 0.10 & 15.22 $\pm$ 0.10 & 15.10 $\pm$ 0.13  \\  
2017-12-09  & 8097.5   & +\,1.3    &           --       &          --      & 15.52 $\pm$ 0.02 & 15.20 $\pm$ 0.03 & 15.06 $\pm$ 0.03  \\  
2017-12-12  & 8100.5   & +\,4.3    &  15.84 $\pm$ 0.12  & 16.12 $\pm$ 0.11 & 15.58 $\pm$ 0.01 & 15.19 $\pm$ 0.04 & 14.99 $\pm$ 0.06  \\  
2017-12-14  & 8102.5   & +\,6.3    &           --       & 16.33 $\pm$ 0.05 & 15.69 $\pm$ 0.04 & 15.19 $\pm$ 0.02 & 14.96 $\pm$ 0.08  \\  
2017-12-17  & 8105.4   & +\,9.2    &  17.00 $\pm$ 0.13  & 16.74 $\pm$ 0.03 & 15.84 $\pm$ 0.05 & 15.32 $\pm$ 0.02 & 15.06 $\pm$ 0.04  \\  
2017-12-26  & 8114.4   & +\,18.2   &  18.50 $\pm$ 0.14  & 17.76 $\pm$ 0.04 & 16.43 $\pm$ 0.02 & 15.81 $\pm$ 0.04 & 15.38 $\pm$ 0.04  \\  
2017-12-29  & 8117.4   & +\,21.2   &           --       &          --      & 16.61 $\pm$ 0.02 & 15.98 $\pm$ 0.02 & 15.59 $\pm$ 0.02  \\  
2018-01-05  & 8124.5   & +\,28.3   &           --       & 18.28 $\pm$ 0.04 & 16.90 $\pm$ 0.06 & 16.27 $\pm$ 0.02 & 15.77 $\pm$ 0.07  \\  
2018-01-07  & 8126.4   & +\,30.2   &  19.07 $\pm$ 0.14  &          --      & 16.96 $\pm$ 0.10 & 16.32 $\pm$ 0.08 & 15.84 $\pm$ 0.07  \\  
2018-01-08  & 8127.4   & +\,31.2   &           --       &          --      & 16.96 $\pm$ 0.06 & 16.25 $\pm$ 0.11 & 15.79 $\pm$ 0.09  \\  
2018-01-14  & 8133.4   & +\,37.2   &           --       & 18.40 $\pm$ 0.03 & 17.14 $\pm$ 0.01 & 16.51 $\pm$ 0.02 & 15.99 $\pm$ 0.03  \\  
2018-01-16  & 8135.4   & +\,39.2   &  19.10 $\pm$ 0.20  & 18.50 $\pm$ 0.04 & 17.21 $\pm$ 0.02 & 16.57 $\pm$ 0.04 & 16.03 $\pm$ 0.06  \\   
2018-01-17  & 8136.5   & +\,40.3   &           --       & 18.52 $\pm$ 0.06 & 17.24 $\pm$ 0.02 & 16.62 $\pm$ 0.02 & 16.05 $\pm$ 0.05  \\   
2018-01-21  & 8140.4   & +\,44.2   &           --       & 18.49 $\pm$ 0.03 & 17.31 $\pm$ 0.02 & 16.72 $\pm$ 0.01 & 16.12 $\pm$ 0.05  \\   
2018-01-30  & 8149.3   & +\,53.1   &           --       & 18.66 $\pm$ 0.02 & 17.40 $\pm$ 0.10 & 16.86 $\pm$ 0.10 & 16.19 $\pm$ 0.04  \\   
2018-02-02  & 8152.4   & +\,56.2   &           --       & 18.57 $\pm$ 0.05 & 17.50 $\pm$ 0.05 & 16.94 $\pm$ 0.03 & 16.34 $\pm$ 0.04  \\   
2018-02-03  & 8153.3   & +\,57.1   &  19.22 $\pm$ 0.21  & 18.61 $\pm$ 0.12 & 17.51 $\pm$ 0.07 & 16.94 $\pm$ 0.06 & 16.32 $\pm$ 0.07  \\   
2018-02-09  & 8159.4   & +\,63.2   &           --       & 18.65 $\pm$ 0.02 & 17.63 $\pm$ 0.02 & 16.98 $\pm$ 0.02 & 16.41 $\pm$ 0.04  \\   
2018-02-13  & 8163.3   & +\,67.1   &  19.08 $\pm$ 0.15  & 18.73 $\pm$ 0.03 & 17.69 $\pm$ 0.02 & 17.12 $\pm$ 0.03 & 16.50 $\pm$ 0.04  \\   
2018-02-16  & 8166.3   & +\,70.1   &  19.14 $\pm$ 0.17  & 18.70 $\pm$ 0.07 & 17.76 $\pm$ 0.02 & 17.20 $\pm$ 0.03 & 16.57 $\pm$ 0.05  \\   
2018-02-20  & 8170.3   & +\,74.1   &           --       & 18.77 $\pm$ 0.02 & 17.83 $\pm$ 0.01 & 17.28 $\pm$ 0.02 & 16.64 $\pm$ 0.04  \\   
2018-02-25  & 8175.2   & +\,79.0   &           --       & 18.79 $\pm$ 0.11 & 17.90 $\pm$ 0.03 & 17.36 $\pm$ 0.03 & 16.72 $\pm$ 0.03  \\   
2018-03-07  & 8185.2   & +\,89.0   &           --       & 18.96 $\pm$ 0.07 & 18.09 $\pm$ 0.02 & 17.53 $\pm$ 0.02 & 16.87 $\pm$ 0.05  \\   
2018-03-15  & 8193.3   & +\,97.1   &           --       &          --      & 18.22 $\pm$ 0.01 & 17.69 $\pm$ 0.01 & 17.05 $\pm$ 0.02  \\   
2018-03-23  & 8201.4   & +\,105.2  &           --       & 19.18 $\pm$ 0.02 & 18.35 $\pm$ 0.02 & 17.86 $\pm$ 0.02 & 17.19 $\pm$ 0.03  \\ 
2018-03-30  & 8208.3   & +\,112.1  &           --       &          --      & 18.47 $\pm$ 0.05 & 17.92 $\pm$ 0.04 & 17.30 $\pm$ 0.03  \\ 
2018-04-02  & 8211.4   & +\,115.2  &           --       & 19.28 $\pm$ 0.09 & 18.50 $\pm$ 0.05 & 18.02 $\pm$ 0.03 & 17.38 $\pm$ 0.05  \\ 
2018-04-07  & 8216.3   & +\,120.1  &           --       &          --      & 18.56 $\pm$ 0.02 & 17.99 $\pm$ 0.02 & 17.38 $\pm$ 0.02  \\ 
2018-04-15  & 8224.3   & +\,128.1  &           --       & 19.50 $\pm$ 0.02 & 18.71 $\pm$ 0.01 & 18.20 $\pm$ 0.01 & 17.60 $\pm$ 0.03  \\ 
2018-04-21  & 8230.2   & +\,134.0  &           --       & 19.64 $\pm$ 0.02 & 18.85 $\pm$ 0.03 & 18.30 $\pm$ 0.01 & 17.76 $\pm$ 0.02  \\ 
2018-04-28  & 8237.4   & +\,141.2  &           --       & 19.79 $\pm$ 0.17 & 18.98 $\pm$ 0.06 & 18.34 $\pm$ 0.04 & 17.86 $\pm$ 0.04  \\ 
2018-05-01  & 8240.3   & +\,144.1  &           --       &          --      & 19.03 $\pm$ 0.04 & 18.47 $\pm$ 0.03 & 17.88 $\pm$ 0.04  \\ 
2018-05-30  & 8269.2   & +\,173.0  &           --       &          --      &         --       & 18.83 $\pm$ 0.02 & 18.38 $\pm$ 0.04  \\ 
2018-07-05  & 8305.2   & +\,209.0  &           --       &          --      & 20.20 $\pm$ 0.03 & 19.41 $\pm$ 0.02 & 18.96 $\pm$ 0.03  \\ 
2018-08-31  & 8362.1   & +\,265.9  &           --       &          --      & 21.22 $\pm$ 0.14 & 20.28 $\pm$ 0.12 & 20.16 $\pm$ 0.15  \\ 
\hline
\end{tabular}\\
$^{\ast}$ With reference to the $B$-band maximum (JD~2458096.2).
\end{table*}

To calibrate a sequence of secondary standards in the SN field, we observed Landolt photometric standard fields \citep{1992AJ....104..340L} PG0231+051, PG0918+029, PG0942-029, and PG1047+003 on three nights, 2018, January 21, February 03 and 25, under good photometric conditions along with the SN field. The observed Landolt field stars cover a brightness range of 12.27 $\le V \le$ 16.11 mag and colour range of $-0.33 \le B-V \le 1.45$ mag. The average atmospheric extinction values in $U$, $B$, $V$, $R$ and $I$ bands for the site were adopted from \citet{2008BASI...36..111S}. 
The average colour terms for the telescope detector system were used to determine the photometric zero-points by applying a linear relationship between the observed and standard colours. A set of 14 stars in the SN field (marked in Fig.~\ref{fig_1}), were calibrated using the estimated zero-points and average colour terms.
The calibrated $UBVRI$ magnitudes of the secondary standards averaged over three nights are listed in Table~\ref{tab_id}.
Considering the proximity of SN to an H {\sc II} region, template subtraction was performed using the template images of the host galaxy, observed on 2019 August 21 with the HCT. The instrumental magnitudes of the supernova at different epochs were extracted from the template subtracted images. The SN magnitude is calibrated differentially with respect to the secondary standards in the SN field by applying the nightly zero points \citep[see][]{2018MNRAS.473.3776K}. The photometric magnitudes of SN along with the errors are listed in Table~\ref{tab_hct}. Here, photometric errors include the errors in aperture photometry as estimated in IRAF and uncertainties in the nightly zero points, added in quadrature.

\subsection{Spectroscopic observation}\label{obs_sp}
Low-resolution optical spectroscopic observations of SN~2017iro were obtained at 34 epochs during 2017 December 01 (JD~2458089.4) to 2018 July 05 (JD~2458305.2). Two grisms Gr\#7 (3500--7800 \AA) and Gr\#8 (5200--9250 \AA) available with HFOSC, were used to obtain the SN~2017iro spectra. 
The spectra of arc lamp and spectrophotometric standards were obtained along with the SN  for calibration purposes. The spectroscopic data reduction was carried out in a standard way using {\small IRAF}. After correcting the observed frames for bias and flat fields, one-dimensional spectra were extracted using the optimal extraction method. Wavelength calibration was done by applying the dispersion solutions obtained using arc lamp spectra. The night-sky emission lines were utilized to secure the wavelength calibration, and small shifts were applied wherever necessary. The spectrophotometric standard observed on the same night was used in obtaining the instrumental response curves for flux calibration of the spectra. For those nights where standard star observations were not possible, the response curves obtained during nearby nights were used. To construct a single flux calibrated spectrum, the spectrum in both the grisms (Gr\#7, Gr\#8) were combined. The spectra were then scaled with respect to the calibrated $UBVRI$ fluxes to bring them to an absolute flux scale. Finally, the SN spectra were corrected for the host galaxy redshift of $z$ = 0.006191 (from NED).

\begin{table*}
\scriptsize
\centering
\caption{Estimated explosion epoch of SN~2017iro using different methods (see Section~\ref{expl}).}
\label{tab_exp}
\small
\begin{tabular}{lc}
\hline
Method adopted                                          &  Explosion epoch                 \\ 
\hline
Power law [$L(t) = A\,\times\,(t - t_{0})^{0.78}$] fit  & JD~2458077.1 (2017 November 19.6)\\
Analytical fit \citep{2018AA...609A.136T}               & JD~2458080.0 (2017 November 22.5)\\
Toy model fit \citep{2008MNRAS.383.1485V}               & JD~2458084.0 (2017 November 26.5)\\
\hline
Mean explosion epoch                                    & JD~2458080.4 (2017 November 22.9)\\    
\hline
\end{tabular}
\end{table*}

\begin{table*}
\scriptsize
\centering
\caption{Light curve parameters of SN~2007iro (see Section~\ref{lc_fit}).}
\label{tab_lc_p}
\small
\begin{tabular}{ccccc|cc|cc} \hline
Filter & Max epoch           & $\rm m_{max}$    & $\rm \Delta m_{15}$ & $\rm \Delta m_{40}$ & \multicolumn{2}{c}{Late time slope} & $\rm \Delta d_{0.25}$\\
       & (JD)                &    (mag)         & (mag)               & (mag)               & ($\sim$40\,--\,140 d)               & ($>$140d)   & (d)    \\ 
       &                     &                  &                     &                     & (mag 100 d$^{-1}$)                  & (mag 100 d$^{-1}$)&  \\ 
\hline 
$U$    & 2458095.7 $\pm$ 0.5 & 15.57 $\pm$ 0.01 & 2.41 $\pm$ 0.07     & 3.55 $\pm$ 0.01     & ---      & ---                      & 7.70 $\pm$ 0.14      \\
$B$    & 2458096.2 $\pm$ 0.5 & 15.94 $\pm$ 0.01 & 1.46 $\pm$ 0.05     & 2.44 $\pm$ 0.01     & 1.27     & ---                      & 10.40 $\pm$ 0.14     \\
$V$    & 2458098.0 $\pm$ 0.5 & 15.55 $\pm$ 0.01 & 0.77 $\pm$ 0.04     & 1.70 $\pm$ 0.01     & 1.73     & 1.80                     & 11.90 $\pm$ 0.14     \\
$R$    & 2458099.4 $\pm$ 0.5 & 15.14 $\pm$ 0.01 & 0.69 $\pm$ 0.02     & 1.54 $\pm$ 0.01     & 1.57     & 1.52                     & 12.20 $\pm$ 0.14     \\
$I$    & 2458100.5 $\pm$ 0.5 & 14.97 $\pm$ 0.01 & 0.52 $\pm$ 0.02     & 1.14 $\pm$ 0.01     & 1.75     & 1.90                     & 14.70 $\pm$ 0.14     \\ 
\hline
Bolometric &  2458095.7 $\pm$ 0.5 &  16.12 $\pm$ 0.01 & 1.05 $\pm$ 0.04 & 1.89 $\pm$ 0.01   & 1.65     & ---                      & 10.40 $\pm$ 0.14     \\ 
\hline
\end{tabular}
\end{table*}

\begin{figure*}
\centering
\resizebox{\hsize}{!}{\includegraphics{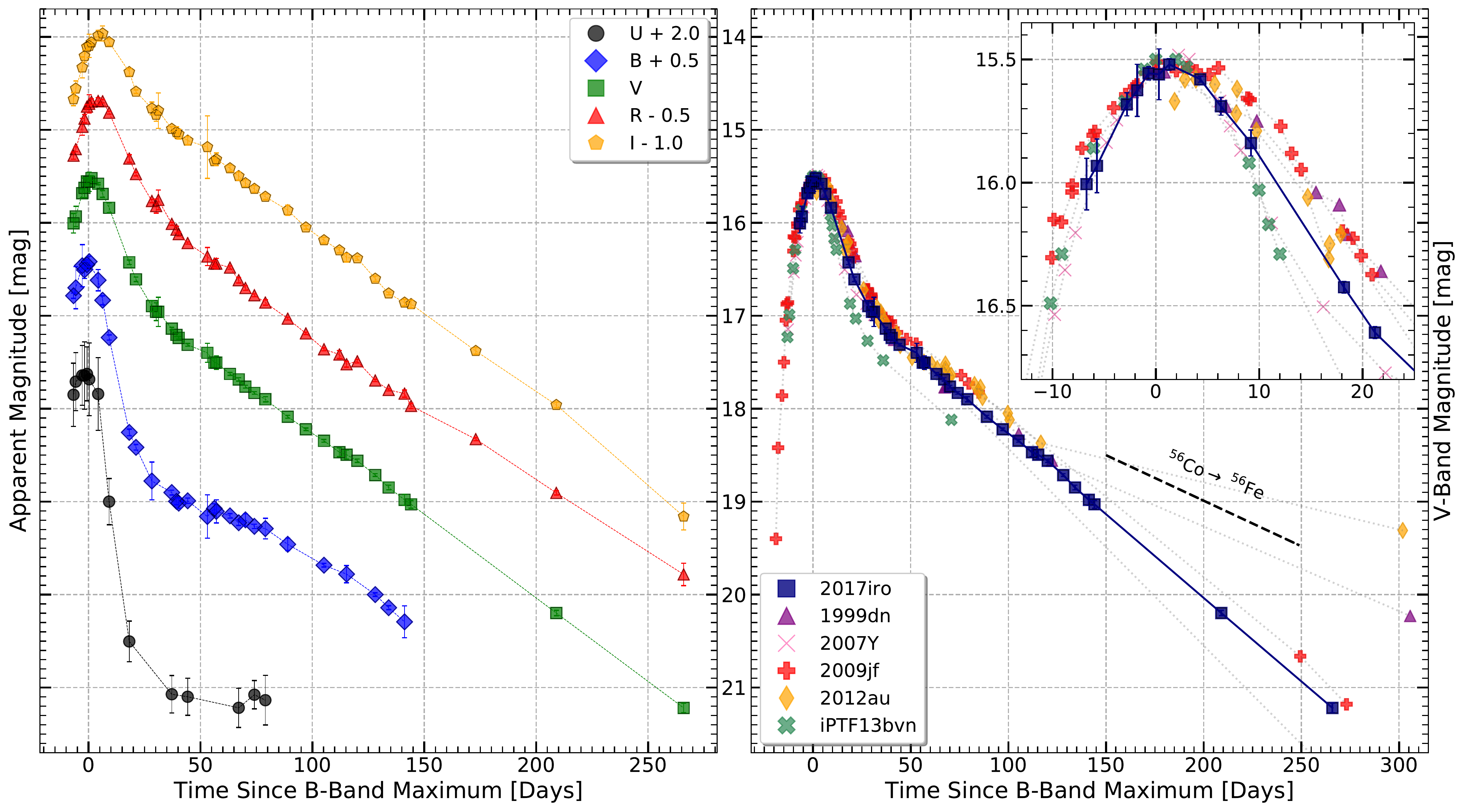}}
\caption{Left panel: The observed $UBVRI$ light curves of SN~2017iro. For clarity, the light curves in different bands have been shifted vertically by the indicated amount. Right panel: The $V$-band magnitude of SN~2017iro is over-plotted with other Type Ib events i.e., iPTF13bvn, SN~2012au, SN~2009jf, SN~2007Y, and SN~1999dn (see Table~\ref{sn_inf} for references). The light curves of other events have been shifted arbitrarily to match the date of maximum and the peak magnitude to SN~2017iro. The evolution of light curves near the peak is also shown in the inset.\label{fig_lc}}
\end{figure*}

\section{Photometric evolution}\label{lc}
\subsection{Estimation of explosion epoch and optical light curves}\label{expl}

The last non-detection of SN~2017iro was reported \citep{2017TNSTR1354....1W} on 2017 November 15.1 (JD~2458072.6) with a limiting magnitude of 19 mag. However, the SN was discovered on 2017 November 30.5 (JD~2458088.0). Since there is a large gap between the first detection and last non-detection various approaches were considered for estimating the explosion epoch. The explosion epoch was estimated by fitting a power law of the form $L(t) = A\,\times\,(t - t_{0})^{0.78}$ \citep{2013ApJ...769...67P} to the pre-maximum phase of the LC. Here, $L$ denotes the luminosity at time $t$, $A$ is a constant that defines rising rate, and $t_{0}$ is the time of the explosion. The best fit value of $t_{0}$ is computed as JD~2458077.1 (2017 November 21.1). While evaluating the explosion parameters of SN~2017iro (see Section~\ref{bol}), the model of \cite{2008MNRAS.383.1485V} was best fitted for an explosion date of JD~2458084.0 (2017 November 26.5). Analytical function \citep{2018AA...609A.136T} fitted to the LC (see Section~\ref{lc_fit}) yielded an explosion epoch of JD~2458080.0 (2017 November 22.5). An average of the above values, i.e., JD~2458080.4\,$\pm$\,2.0 (2017 November 22.9), was adopted as the explosion epoch of SN~2017iro (see Table~\ref{tab_exp}). Further, the typical rise times of Type Ib SNe in $R$-band and bolometric light curve are $\sim$\,22 d \citep{2015A&A...574A..60T} and $\sim$17 d \citep{2016MNRAS.457..328L}, respectively. The $R$-band and the bolometric peak of SN~2017iro occurred on JD~2458099.4 and JD~2458096.6, respectively (see Section~\ref{lc_fit}). The mean explosion date estimated above is consistent with the rise time criteria.

The $UBVRI$ light curves of SN~2017iro are shown in Fig.~\ref{fig_lc} (left panel). A well-sampled light curve covering the pre-maximum to nebular phase was obtained using the photometric observation made on 40 epochs. In the right panel of Fig.~\ref{fig_lc}, the $V$-band light curve of well-studied Type Ib SNe: iPTF13bvn \citep{2014MNRAS.445.1932S,2015A&A...579A..95K}, SN~2012au \citep{Pandey_12au}, SN~2009jf \citep{2011MNRAS.413.2583S,2011MNRAS.416.3138V}, SN~2007Y \citep{2009ApJ...696..713S}, and SN~1999dn \citep{2011MNRAS.411.2726B}, having observations during pre-maximum and nebular phases are plotted along with SN~2017iro for comparison. The light curve evolution and estimation of related parameters are discussed in the following subsections. 

\subsection{Light curve parameters}\label{lc_fit}

Various light curve parameters were obtained by fitting the light curves in different bands with an analytical function proposed by \citet[][see their section 3]{2018AA...609A.136T}. This method is based on the empirical model suggested by \citet{1996ApJ...471L..37V} for thermonuclear events whose LCs are similar to that of SE-SNe. The epoch of maximum and the light curve parameters of SN~2017iro estimated from the LCs of different pass-bands are listed in Table~\ref{tab_lc_p} and are compared with other SE-SNe. A lag of $\sim$5 d between $U$ and $I$-band maximum is inferred for SN~2017iro. This duration is shorter in comparison to $\sim$8 d for the slow-declining SN~2009jf \citep{2011MNRAS.413.2583S} and $\sim$9 d for the fast-declining iPTF13bvn \citep{2014MNRAS.445.1932S}. The early appearance of maximum in bluer bands is a common behaviour of SE-SNe, and SN~2017iro follows this trend.

The luminosity decline rate parameter, $\rm \Delta m_{15}$ (difference in magnitude at peak and 15 d later), is handy in the context of Type Ia SNe, where it is found to be correlated with the peak luminosity. A smaller $\rm \Delta m_{15}$ value favours more luminous objects \citep{1993ApJ...413L.105P}. Large data sets obtained from different surveys have provided an opportunity to investigate if such a correlation exists for SE-SNe. No direct correlation was found between the light curve decline rate parameter and luminosity by \citet{2011ApJ...741...97D} and \citet{2015A&A...574A..60T}. However, the sample study of a larger set of SE-SNe in \citet{2018AA...609A.136T} hinted towards the fact that luminous events tend to have broader light curves, i.e., a smaller value of $\rm \Delta m_{15}$. In the following paragraph, we compared various light curve parameters of SN~2017iro with some well-studied Type Ib SNe.

The $\rm \Delta m_{15}$ values of SN~2017iro are 1.46, 0.77, 0.69 and 0.52 mag in $B$, $V$, $R$ and $I$ bands, respectively. These are similar to the values inferred for SN~2012au \citep[1.60, 0.90, 0.70, and 0.43 mag, respectively]{Pandey_12au}. On the other hand, iPTF13bvn shows comparatively rapid evolution \citep[1.82, 1.16, 1.22 and 0.90 mag, respectively,][]{2014MNRAS.445.1932S} whereas SN~2009jf shows a relatively slower evolution \citep[0.91, 0.50, 0.31, and 0.31 mag, respectively]{2011MNRAS.413.2583S}. The above can be inferred from the comparison of normalised $V$-band light curves shown in right panel of Fig.~\ref{fig_lc}. The comparison of early phase light curve indicates that the decline rate of SN~2017iro is intermediate between the fast declining SN~2007Y, iPTF13bvn and the slow declining SN~2009jf. The intermediate width of light curves of SN~2017iro could possibly indicate that ejecta mass of SN~2017iro lies in between the inferred ejecta mass of iPTF13bvn and SN~2009jf (see Section~\ref{exp_pa}).

The decay in the magnitude from the peak to +40 d is denoted by the $\rm \Delta m_{40}$ parameter by \citet{2018AA...609A.136T}. $\rm \Delta m_{40}$ was found to be in the range of 1.5 to 1.7 mag (in $r$-band) for their sample of SE-SNe. SN~2017iro lies towards the lower end (1.54 mag in $R$-band) of the observed range, indicating a relatively faster evolution for a SE-SN. The late phase ($>$\,40 d) light curve decay rate of SN~2017iro is also computed in all bands (see Table~\ref{tab_lc_p}). The decay rates between $\sim$\,+40 to $\sim$\,+140 d are 1.27, 1.73, 1.57 and 1.75 mag 100 d$^{-1}$ (in $B$, $V$, $R$ and $I$ bands, respectively) and beyond $\sim$140 d is 1.80, 1.52 and 1.90 mag 100 d$^{-1}$ (in $V$, $R$ and $I$ bands, respectively). It is to be noted that usually, SE-SNe display faster decline than Type II events during the late phases. Such features are indicative of higher $\gamma$-rays escape. The implications of the fast declining nature of SN~2017iro is discussed in Section~\ref{late_lc}.

The $V$-band absolute magnitude ($M_V$) of SE-SNe is known to display a wide range between $\sim$\,--16.5 to $\sim$\,--19.5 mag \citep{2006AJ....131.2233R, 2011ApJ...741...97D, 2018AA...609A.136T}. To estimate the $M_V$ of SN~2017iro, we adopted distance of the host galaxy NGC~5480 as 30.8\,$\pm$\,2.2 Mpc (corrected for Virgo infall only) and the corresponding distance modulus 32.44\,$\pm$\,0.15 mag (for an H$_{0}$ = 73 km s$^{-1}$ Mpc$^{-1}$). The adopted distance is consistent with the distance estimate of 30.73 Mpc using Tully-Fisher relation \citep{2007-Theureau}. After applying a correction for the reddening $E(B-V)$ = 0.28 mag (see Section~\ref{color}), the $M_V$ of SN~2017iro is derived as $-17.76 \pm 0.15$ mag. This indicates that SN~2017iro is fainter than SN~2007uy, SN~2009jf and SN~2012au but brighter than SN~2007Y and iPTF13bvn.

\begin{figure}
\centering
\includegraphics[width=\columnwidth]{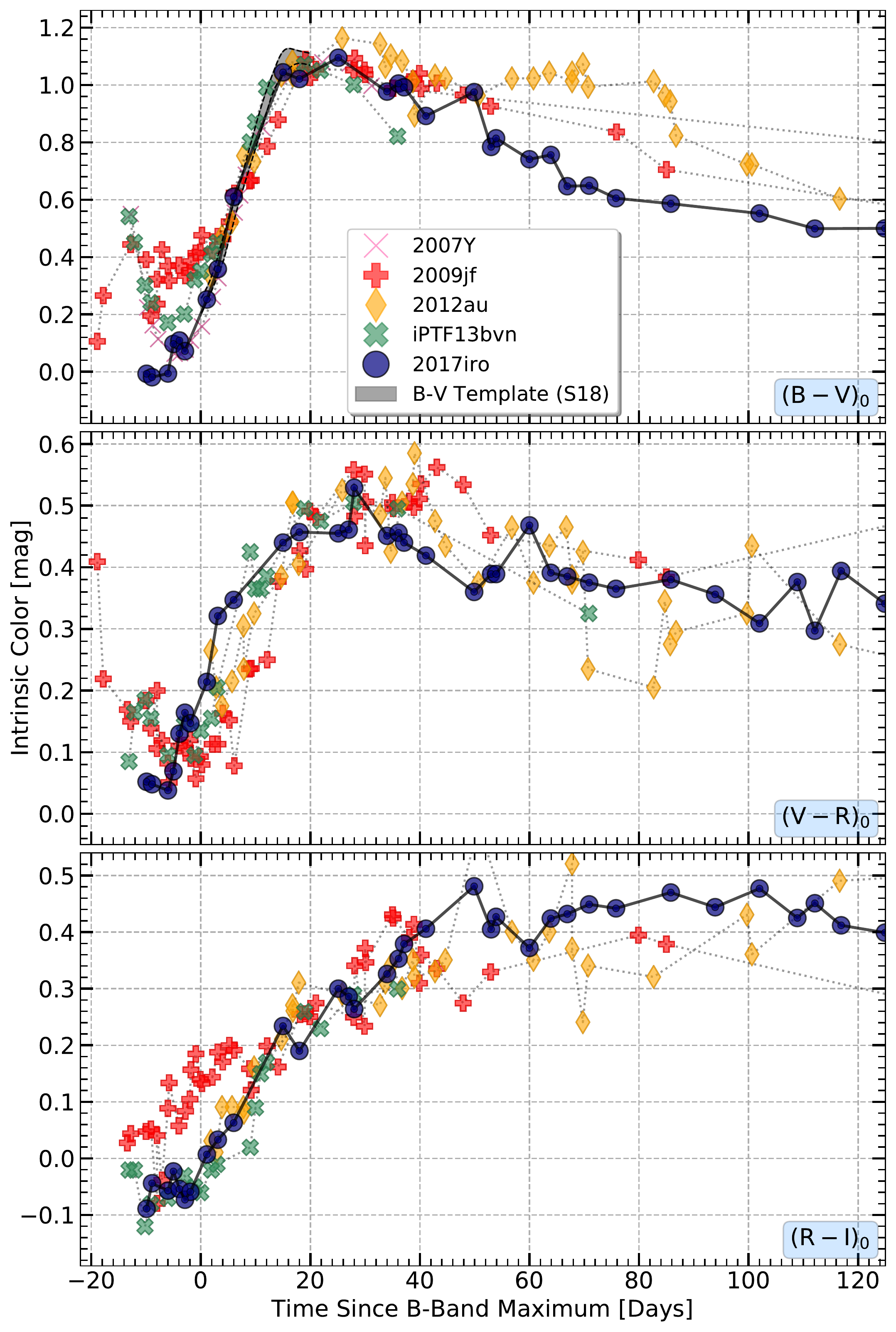}
\caption{Colour evolution of SN~2017iro, plotted along with iPTF13bvn, SN~2012au, SN~2009jf, SN~2007uy and SN~2007Y. Both the MW and the host galaxy extinction corrections have been applied for each SN. The $B-V$ colour templates from \citet{2018AA...609A.135S} are over-plotted in the top panel for comparison. The bibliographic sources are the same as mentioned in the text (Section~\ref{lc}).}
\label{fig_col}
\end{figure}

\begin{figure}
\centering
\includegraphics[width=\columnwidth]{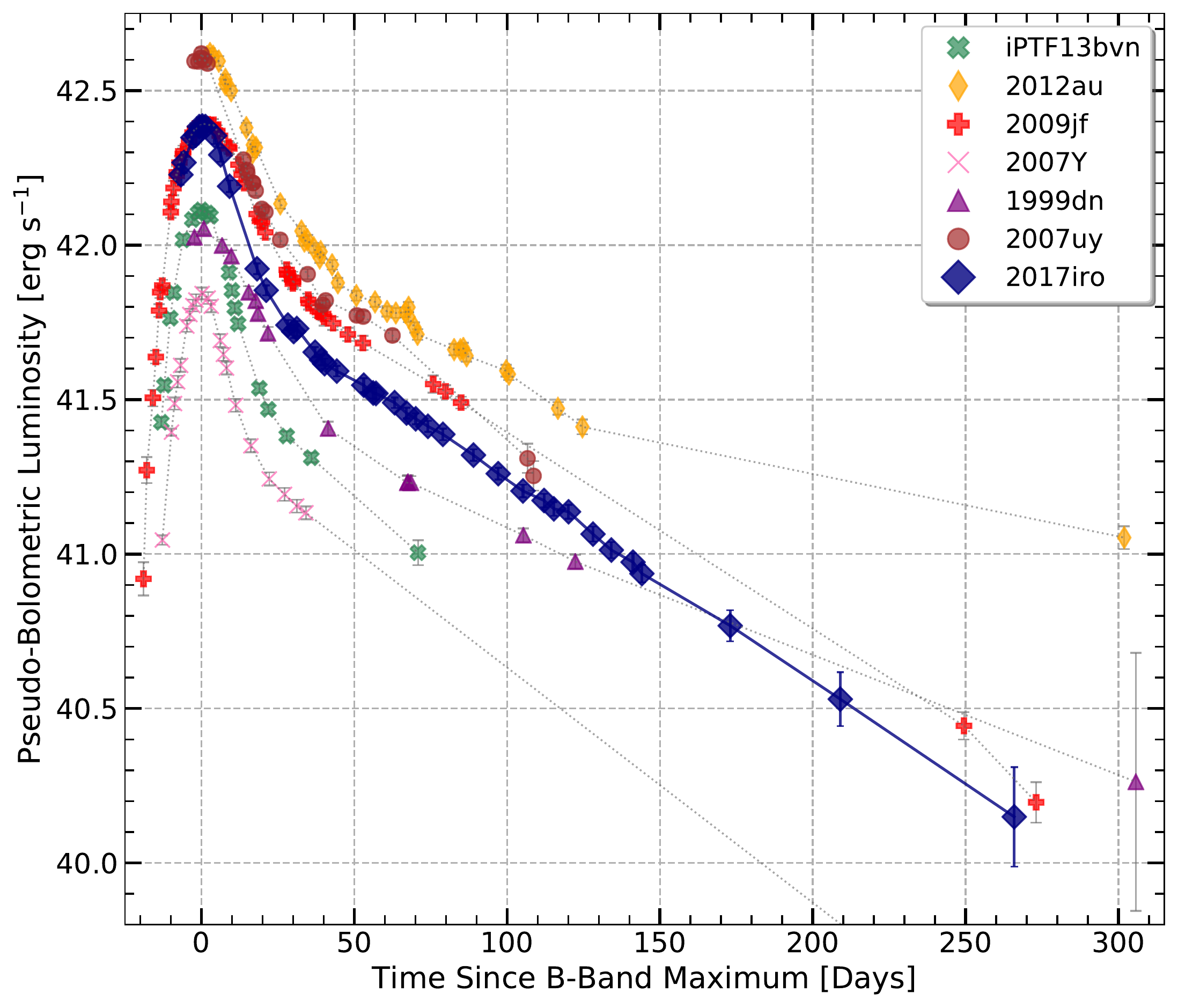}
\caption{The psuedo-bolometric light curve of SN~2017iro (connected with a bold line) compared with other similar Type Ib events (iPTF13bvn, SN~2012au, SN~2009jf, SN~2007uy, SN~2007Y and SN~1999dn). The psuedo-bolometric light curves were obtained using \textsc{SUPERBOL}.}
\label{bol_mag}
\end{figure}

\subsection{Reddening and colour evolution}\label{color}

The estimate of reddening suffered by SN is crucial as it affects the explosion parameters which are derived using the observables. The reddening due to ISM in the Milky Way along the line of sight (LOS) of the SN can be inferred using the all-sky dust extinction map \citep[e.g.][]{2011ApJ...737..103S, 2016ApJ...821...78S}. The determination of reddening within the host galaxy is a challenging task. In general, the sites of SE-SNe, are associated with star-forming regions \citep[e.g.][]{2012MNRAS.424.1372A, 2013MNRAS.436..774E} and may suffer a considerable amount of reddening. Several photometric and spectroscopic techniques have been proposed to determine the reddening within the host galaxy but with caveats. 

The Galactic reddening along the LOS of SN~2017iro as derived from the all-sky dust extinction map \citep*{2011ApJ...737..103S} is $E(B-V)$ = 0.016 $\pm$ 0.001 mag. The equivalent width (EW) of Na\,{\sc i} D absorption lines is considered as a good tracer of reddening within the host-galaxy \citep{1990-Barbon, 2003fthp.conf..200T, 2012MNRAS.426.1465P} but with caveats \citep{2011MNRAS.415L..81P, 2013ApJ...779...38P}. To estimate the EW of Na\,{\sc i} D, we stacked the low-resolution spectra of SN~2017iro obtained near the maximum light. A reasonably strong Na\,{\sc i} D absorption feature is seen at the redshift of the host galaxy with an EW\,=\,1.34\,$\pm$\,0.10\,\AA. An average value of $E(B-V)_{host}$, computed from the relation of \citet{1990-Barbon} and \citet{2003fthp.conf..200T} is 0.27\,$\pm$\,0.04 mag. For further analysis we adopt a total extinction $E(B-V)_{MW+host}$ = 0.28\,$\pm$\,0.04 mag (Milky Way + host) and \citet{1989-Cardelli} extinction law with ratio of selective to total extinction $R_{V}$\,=3.1.

In Fig.~\ref{fig_col}, the $B-V$, $V-R$, and $R-I$ colours of SN~2017iro are compared with other well-studied Type Ib events (see Table~\ref{sn_inf} for details and references). It is seen that the overall shape of colour evolution for each colour is similar \citep[also see][]{2018AA...609A.135S}. Up to $\sim$\,+20 d, the $B-V$ colour evolves faster than the $V-R$ and $R-I$ colours. During the late phase ($>$\,20 d), the colour evolution is almost flat.

\begin{table}
\scriptsize
\centering
\caption{Sample of Type Ib SNe used in this study.}
\begin{tabular}{lccc} \hline
Name      &  $E(B-V)_{tot}$  & Distance     & References \\
          &    (mag)         & (Mpc)        &            \\ \hline
SN~1999dn &     0.10         &   38.90      & 1          \\
SN~2007Y  &     0.11         &   19.31      & 2          \\
SN~2007uy &     0.63         &   29.50      & 3          \\
SN~2009jf &     0.11         &   34.25      & 4, 5       \\
SN~2012au &     0.06         &   23.50      & 6, 7       \\
iPTF13bvn &     0.22         &   22.49      & 8, 9       \\
\hline
\end{tabular}\\
References:
$^{1}$\,\citep{2011MNRAS.411.2726B},
$^{2}$\,\citep{2009ApJ...696..713S},
$^{3}$\,\citep{2013MNRAS.434.2032R},
$^{4}$\,\citep{2011MNRAS.413.2583S},
$^{5}$\,\citep{2011MNRAS.416.3138V},
$^{6}$\,\citep{2013ApJ...770L..38M},
$^{7}$\,\citep{2013ApJ...772L..17T},
$^{8}$\,\citep{2014MNRAS.445.1932S},
$^{9}$\,\citep{2015A&A...579A..95K}.
\label{sn_inf}
\end{table}

\begin{figure}
\centering
\includegraphics[width=\columnwidth]{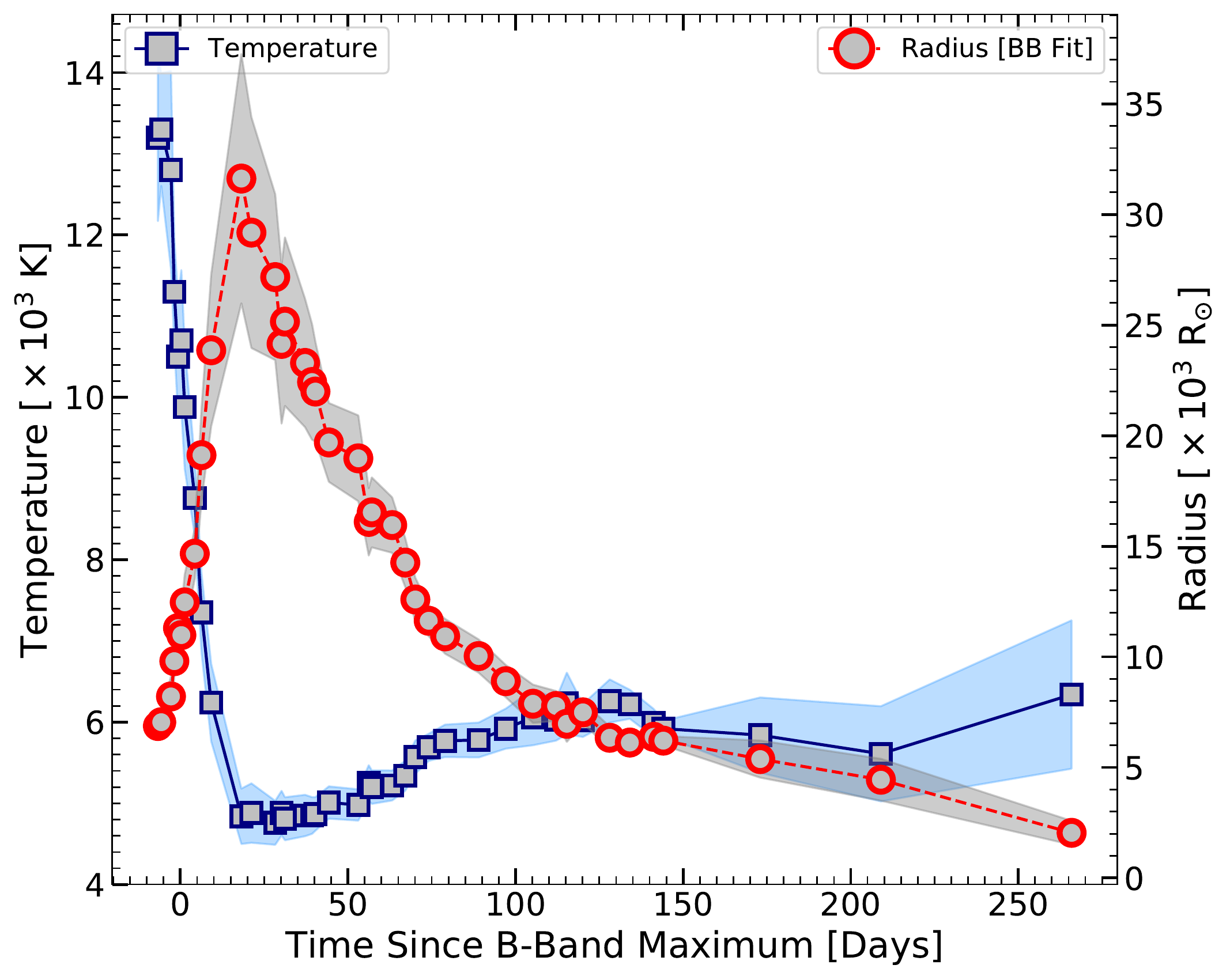}
\caption{The blackbody temperature and radius evolution of SN~2017iro. These parameters were derived from the photometric data using \textsc{SUPERBOL}. The shaded regions indicate 1-$\sigma$ uncertainties.}
\label{BB}
\end{figure}

\begin{figure*}
\centering
\includegraphics[scale = 0.6]{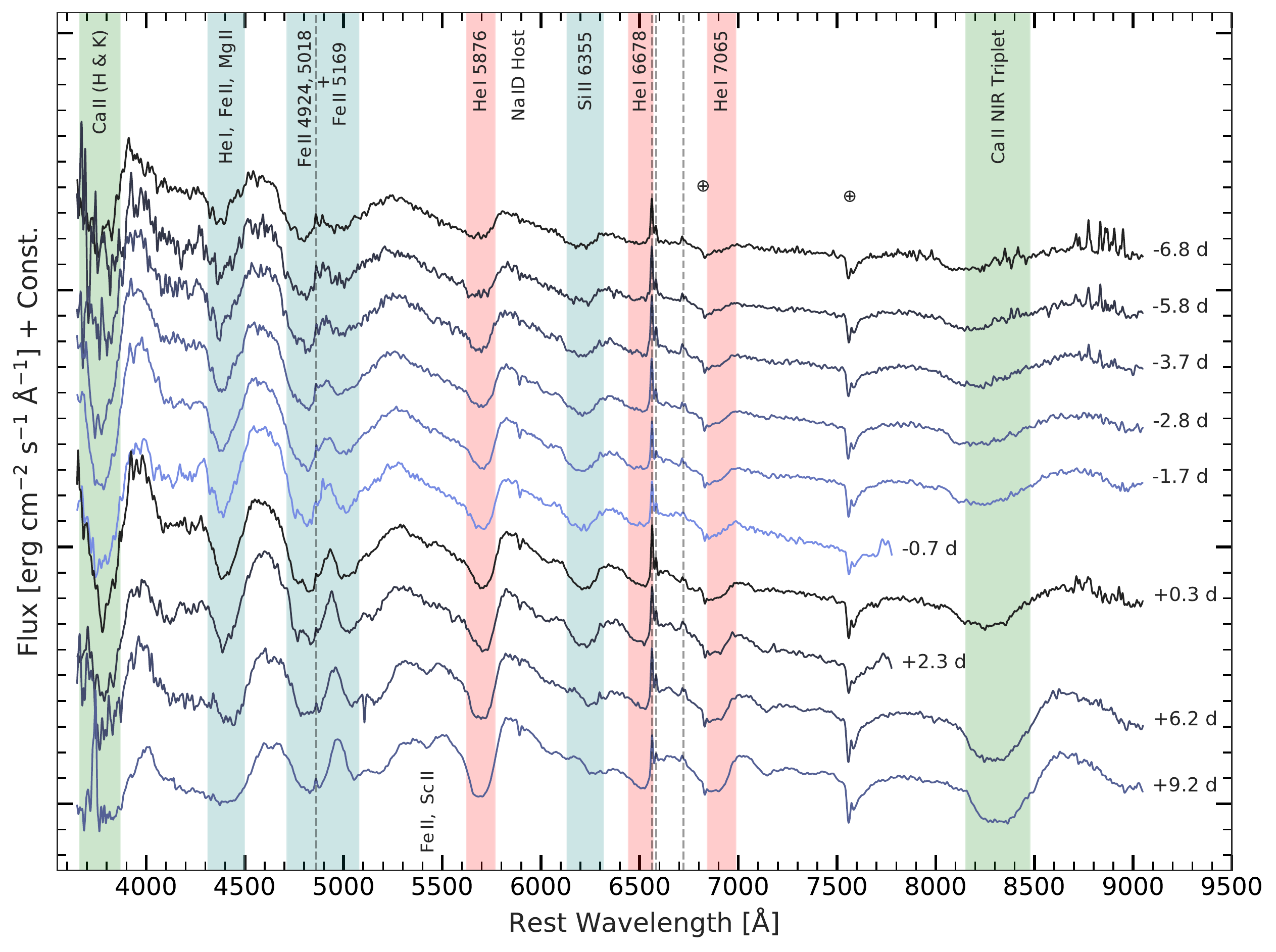}
\includegraphics[scale = 0.6]{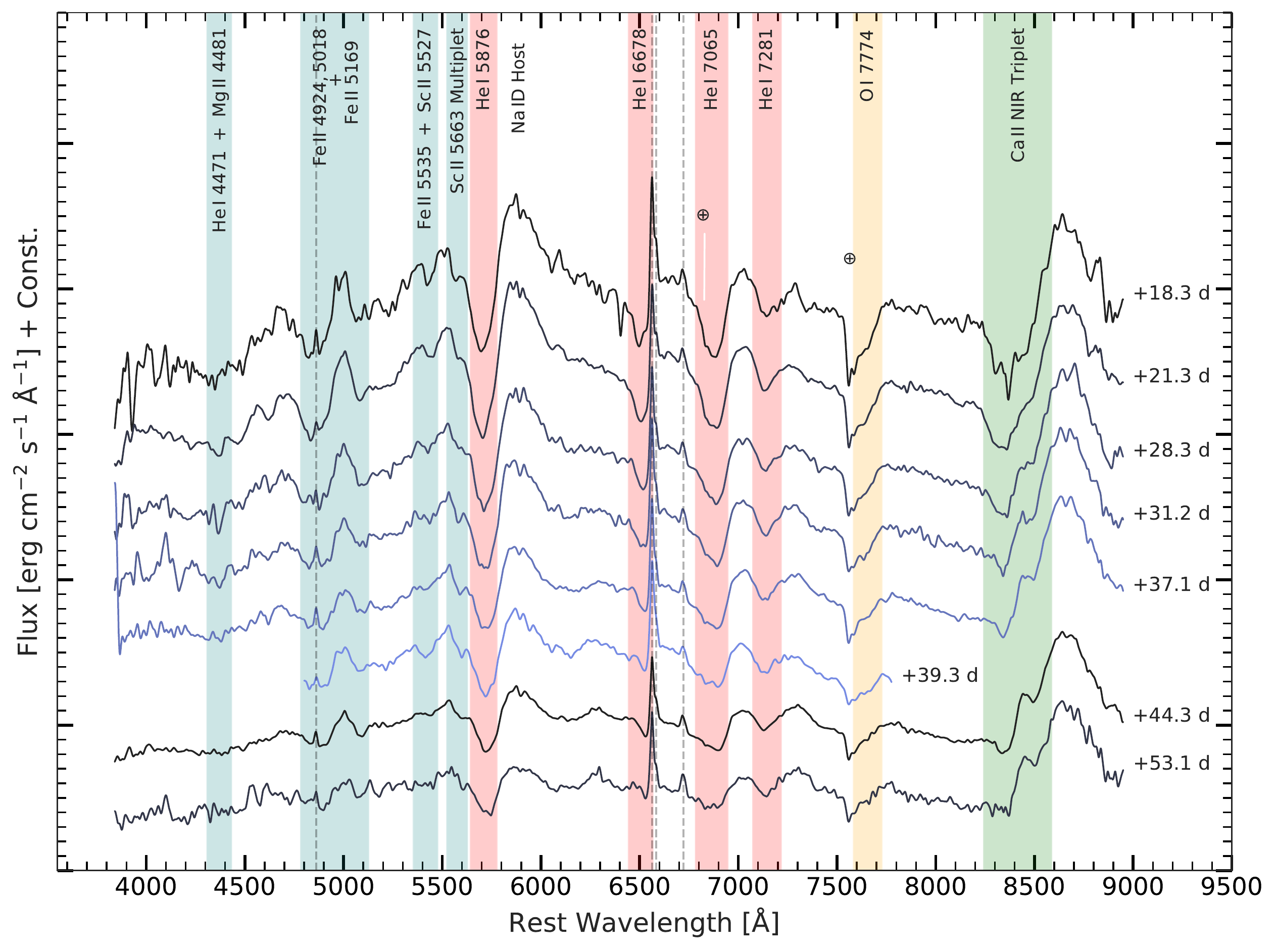}
\caption{Pre-maximum along with early post-maximum (--7 to +9 d) and late post-maximum (+18 to +53 d) phases spectral evolution of SN~2017iro are shown in the top and bottom panels, respectively. Prominent spectral lines are marked. The labelled phases are with respect to the $B$-band maximum (JD~2458096.2). The spectra have been corrected for redshift and reddening. Features arising from the host are marked with dashed vertical lines. Major telluric bands are shown with a circled plus symbol. The spectra shown in Figures \ref{spec_pre}, \ref{fig_neb} and \ref{fig_neb_late} are available as data behind the figure.} \label{spec_pre}
\end{figure*}

\begin{figure}
\centering
\includegraphics[width=\columnwidth]{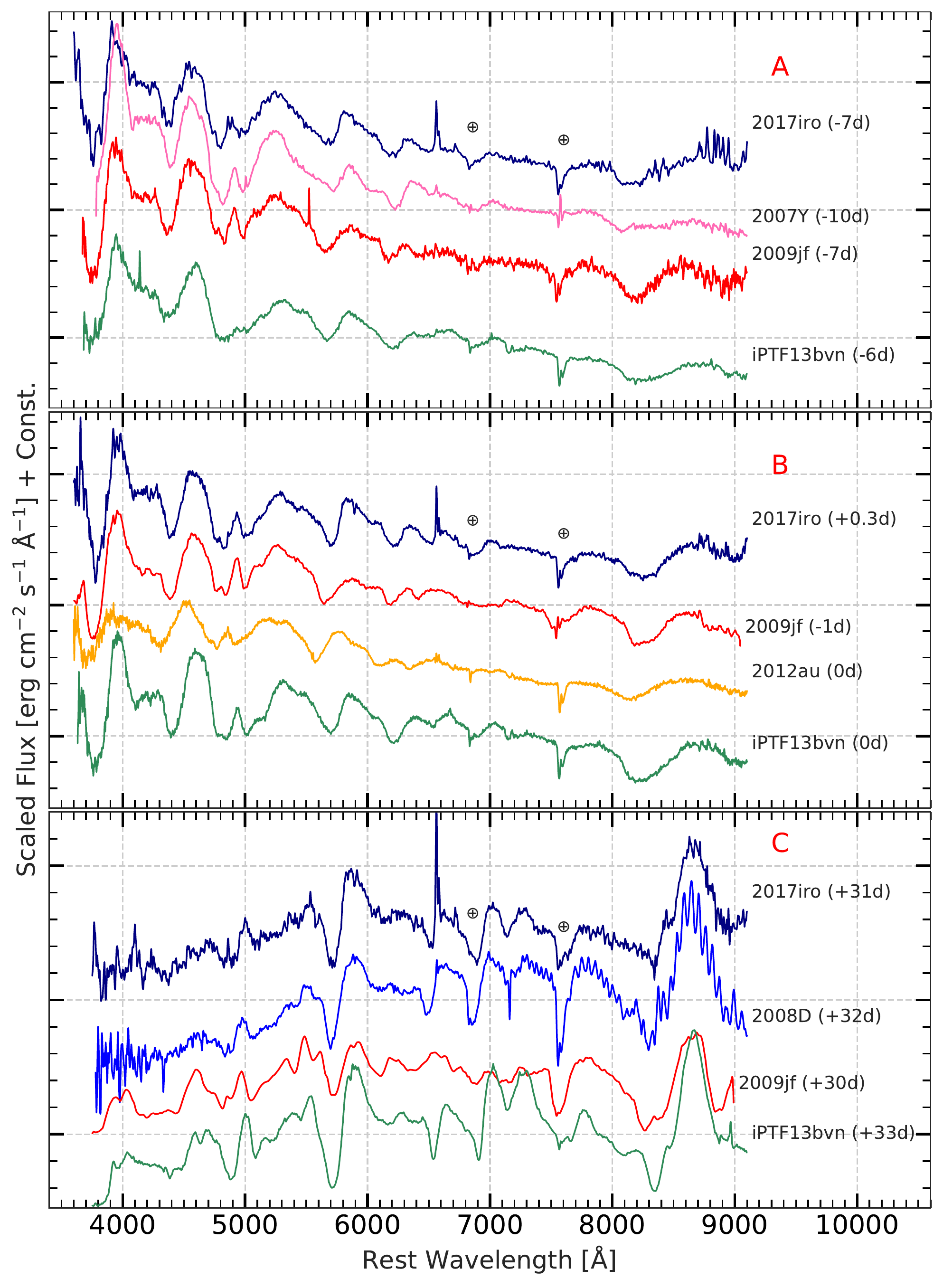}
\caption{SN~2017iro spectral evolution compared with other Type Ib events (iPTF13bvn, SN~2012au, SN~2009jf, SN~2008D and SN~2007uy) at similar epochs. Panels A, B and C represent the pre-maximum, near-maximum and post-maximum spectra, respectively. Telluric lines are shown with circled plus symbols.}
\label{fig_comp1}
\end{figure}

\subsection{Quasi-bolometric light curve}\label{bol}

The quasi-bolometric light curve of SN~2017iro was constructed using the latest version of Python-based code \textit{SuperBol} \citep{2018-Nicholl}. The $UBVRI$ magnitudes after correcting for the total extinction $E(B-V)$ = 0.28 mag, and adopted distance to NGC~5480 were given as input parameters. The $U$ and $B$ band magnitudes were extrapolated by assuming a constant colour during the late phases. The flux integration was performed only over the optical wavelengths. The resultant quasi-bolometric light curve of SN~2017iro along with well studied Type Ib events (estimated similarly) are shown in Fig.~\ref{bol_mag}. Furthermore, the black body fit parameters (temperature and radius) which is a default outcome of the code is also plotted in Fig.~\ref{BB}.

We fitted \citet{2018AA...609A.136T} equation to the quasi-bolometric light curve of SN~2017iro and estimated parameters listed in Table~\ref{tab_lc_p}. The light curve peaked at JD~2458095.7 with log$_{10}L$\,=\,42.39\,$\pm$\, 0.09 erg s$^{-1}$. The $\rm \Delta m_{15_{bol}}$ value of SN~2017iro is 1.05 mag, which is higher than that of SN~2009jf (0.60 mag) but is comparable to SN~2007Y (0.80 mag). The decay rates after +40 d of SN~2017iro (0.016 mag d$^{-1}$) and SN~2009jf (0.014 mag d$^{-1}$) are similar implying relatively faster evolution than the cobalt decay rate ($\sim$0.01 mag d$^{-1}$, see also Section~\ref{late_lc}). This decay rate is consistent with the range 0.014\,--\,0.018 mag d$^{-1}$ seen in Type Ib events \citep{2018AA...609A.136T}. 

\begin{figure*}
\centering
\includegraphics[scale = 0.7]{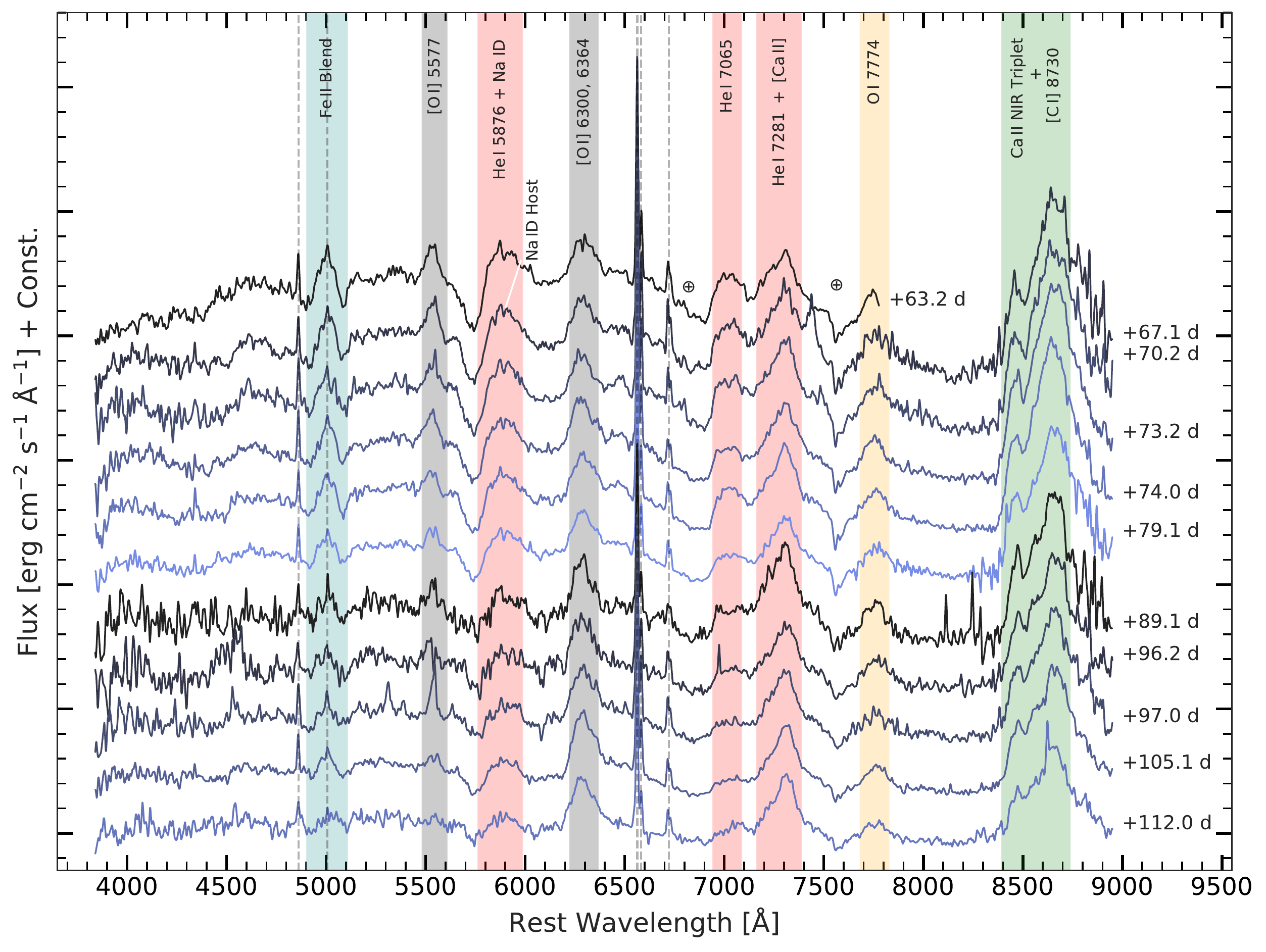}
\caption{Late post-maximum and early nebular phase ($\sim$+67 to +112 d) spectral evolution of SN~2017iro. Prominent spectral lines are also marked. Other descriptions are similar as mentioned in Fig.~\ref{spec_pre}. The spectra shown in Figures \ref{spec_pre}, \ref{fig_neb} and \ref{fig_neb_late} are available as data behind the figure.}
\label{fig_neb}
\end{figure*}

\begin{figure*}
\centering
\includegraphics[scale = 0.7]{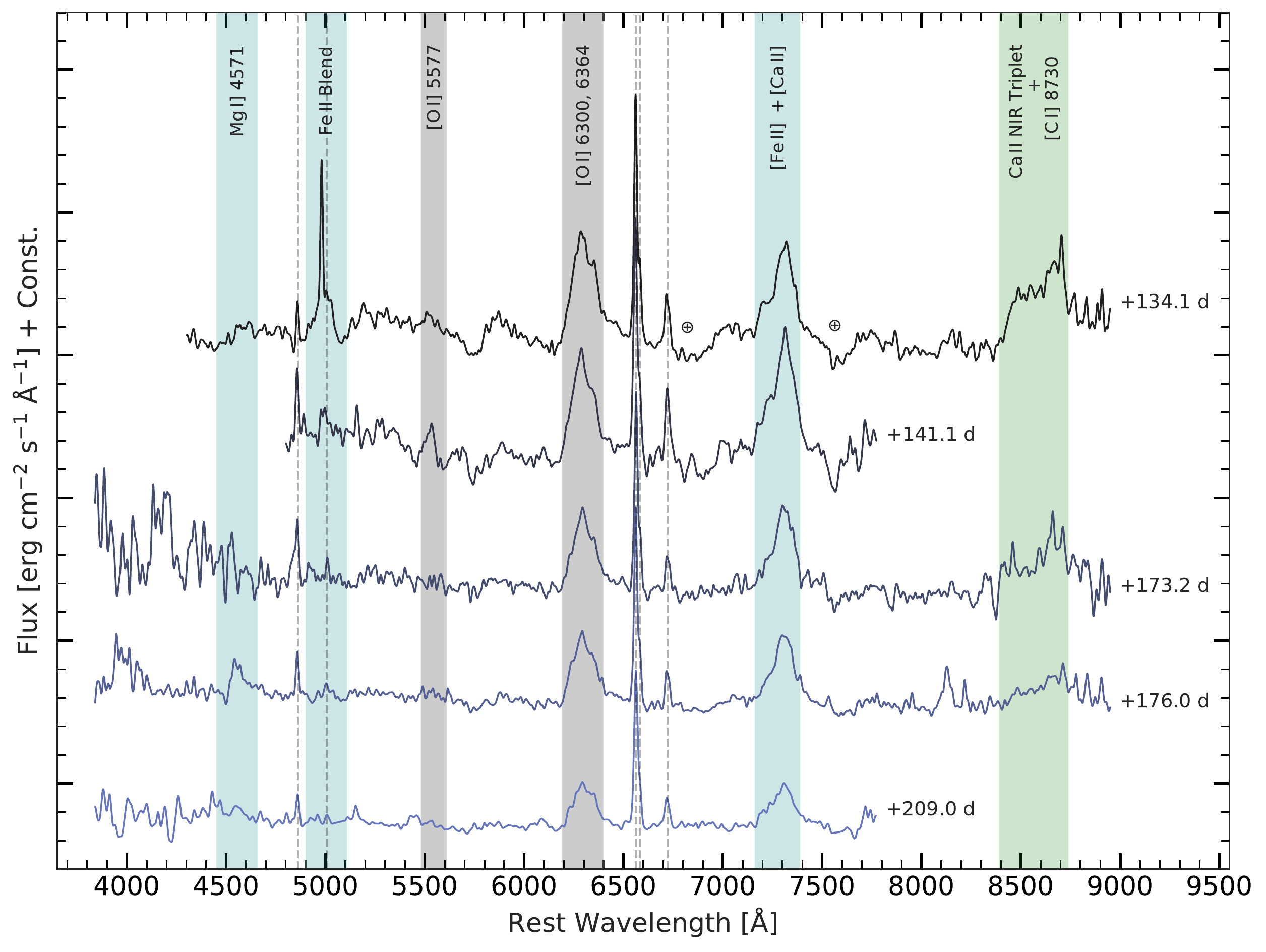}
\caption{Nebular phase ($\sim$+134 to +209 d) spectra of SN~2017iro. Various nebular emission lines are shown. The dashed vertical lines indicate the features arising from the host and the telluric bands are shown with a circled plus symbol (for other descriptions, see Fig.~\ref{spec_pre}). The spectra shown in Figures \ref{spec_pre}, \ref{fig_neb} and \ref{fig_neb_late} are available as data behind the figure.}
\label{fig_neb_late}
\end{figure*}

\begin{figure}
\centering
\includegraphics[width=\columnwidth]{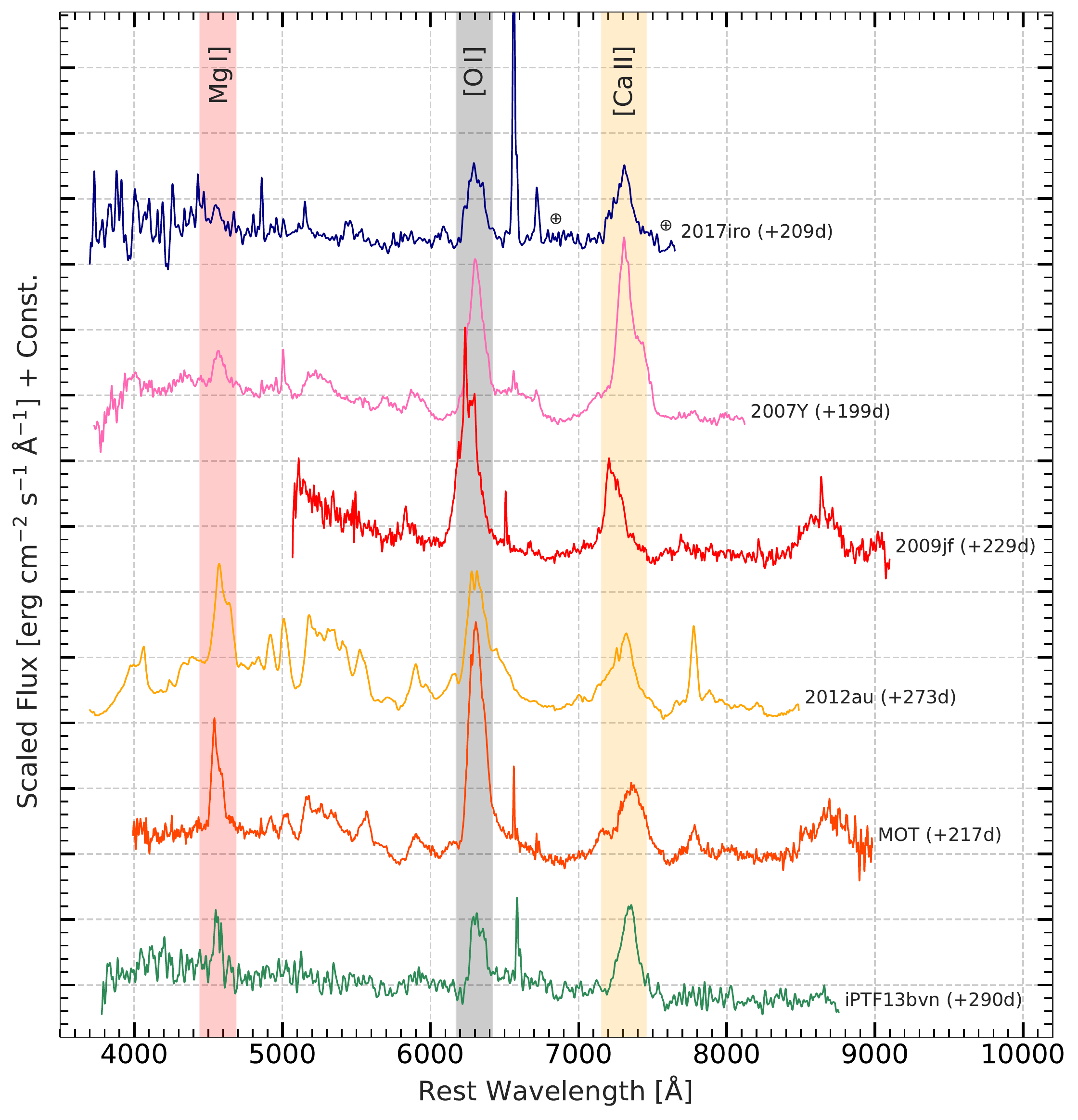}
\caption{Comparison of nebular phase spectral features of SN~2017iro with other well studied Type Ib events. The SN references are same as described in Section~\ref{pre_max_spc} in addition to Master OT J120451.50+265946.6 \citep[MOT,][]{2019MNRAS.485.5438S}.}
\label{fig_nebular}
\end{figure}

\begin{figure*}
\centering
\resizebox{\hsize}{!}{\includegraphics{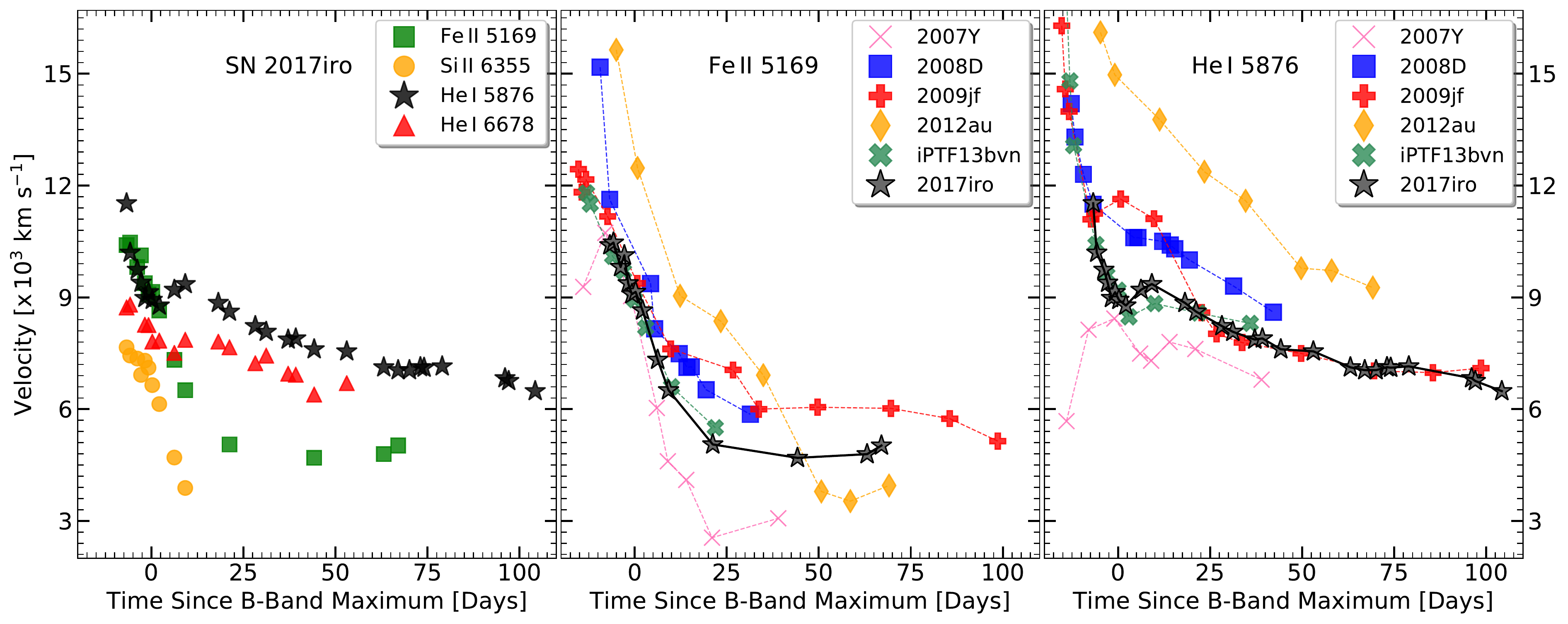}}
\caption{Evolution of the velocity of different spectral lines observed in SN~2017iro. Error in these estimates was never below 800 km s$^{-1}$. The line velocities of different SNe (SN~2007Y, 2008D, 2009jf, 2012au and iPTF13bvn) are also compared.}
\label{fig_c2}
\end{figure*}

\section{Spectroscopic evolution}\label{spec_anal}

The early (photospheric phase) spectra are useful to examine the outermost regions of the expanding ejecta. For example, the presence/absence of residual hydrogen and/or helium during the pre-explosion stellar evolution can be explored with the early phase spectra. In the nebular phase, because of expansion, the ejecta becomes optically thin, and hence spectra during this phase provide essential information about the inner layers of the ejecta.   

In this section, the spectral evolution of SN~2017iro covering pre-maximum to nebular phase is presented. The densely sampled spectroscopic data provided an opportunity to compare important spectral features with other well-studied events. All spectra were corrected for the heliocentric velocity of 1856 km s$^{-1}$ of the host galaxy (NGC~5480). The evolution of spectroscopic features is discussed in the following sections.


\subsection{Pre-maximum and early post-maximum phase}\label{pre_max_spc}

The spectral evolution in the pre-maximum and around maximum phase is shown in Fig.~\ref{spec_pre} (top panel). During early phase (i.e. pre-maximum), spectroscopy of SN~2017iro was done almost every night. The first spectrum at $\sim$--7 d clearly shows the broad absorption feature of He\,{\sc i} 5876 \AA\ line, and weak signatures of other He\,{\sc i} lines. He\,{\sc i} 6678 \AA\ is blended with narrow H${\alpha}$ line from the host galaxy, whereas the He\,{\sc i} 7065 \AA\ feature lies at the edge of the telluric absorption band at 6870 \AA. Other narrow lines (H${\beta}$ 4861 \AA, [N\,{\sc ii}] 6584 \AA\ and [S\,{\sc ii}] 6717, 6731 \AA) originating from the underlying H\,{\sc ii} region are also clearly visible. In the extreme blue region of the observed spectra, well developed Ca\,{\sc ii} H\,\&\,K lines (3934 and 3968 \AA) are seen. The spectrum also shows features of Fe\,{\sc ii} triplet (4924, 5018 and 5169 \AA), Si\,{\sc ii} 6355 \AA\ and a weak absorption due to Ca\,{\sc ii} NIR triplet (8498, 8542 and 8662 \AA).

In the pre-maximum spectral evolution, the region between 4000 to 4300 \AA\ appears flat, possibly due to the blending of metal lines in the spectra. The absorption around 4400 \AA\, is possibly a blend of He\,{\sc i}, Fe\,{\sc ii} and Mg\,{\sc ii}. As the SN evolves towards the maximum, the lines due to He\,{\sc i}, Fe\,{\sc ii}, and Si\,{\sc ii} become more prominent. In the early post-maximum spectra, the Si\,{\sc ii} 6355 \AA\ feature starts weakening and vanishes in the spectra obtained on +18 d (cf. Fig.~\ref{spec_pre}, bottom panel). A feature around 5500 \AA\ begins to develop in the spectrum of $\sim$+6 d, which is likely a blend of Fe\,{\sc ii} 5535 \AA\ and Sc\,{\sc ii} 5527 \AA\ lines. The Na\,{\sc i}\,D lines from the host galaxy are seen as a narrow absorption in the spectra.

The spectra of SN~2017iro during the pre-maximum (Fig.~\ref{fig_comp1}, Panel: A) and close to maximum phase (Fig.~\ref{fig_comp1}, Panel: B) are compared with the spectra of few other well-studied Type Ib SNe SN~2009jf \citep{2011MNRAS.413.2583S}, iPTF13bvn \citep{2014MNRAS.445.1932S,2015A&A...579A..95K}, SN~2007Y \citep{2009ApJ...696..713S}, SN~2008D \citep{2009ApJ...702..226M} and SN~2012au \citep{Pandey_12au} around similar epochs. A good similarity can be noticed in the spectrum of SN~2017iro obtained at --7 d and near maximum with those of SN~2009jf and iPTF13bvn. Except for iPTF13bvn, the Fe\,{\sc ii} line near 5000 \AA\ is found to be prominent in other SNe used in comparison (cf. panel: A in Fig.~\ref{fig_comp1}). However, in SN~2017iro, the Fe\,{\sc ii} line appears to be contaminated by narrow H${\beta}$ line from the underlying H\,{\sc ii} region. The spectrum of SN~2012au is shifted  blue-wards (cf. panel: B in Fig.~\ref{fig_comp1}) due to its higher expansion velocity \citep{2013ApJ...770L..38M,2013ApJ...772L..17T}.

\subsection{Late post-maximum and early nebular phase}\label{post_pk}

The spectral evolution during +18 to +53 d is displayed in the bottom panel of Fig.~\ref{spec_pre}. The continuum becomes redder during this phase, and the Ca\,{\sc ii} H\,\&\,K lines disappear, and other metallic lines appear in this part of the spectrum. However, the region below 4500 \AA\, in our spectrum, is noisy, so lines could not be identified. The He\,{\sc i} lines are initially prominent, but beyond +45 d, they weaken. In the blue wing of He\,{\sc i} 5876 \AA\, line, lines due to Sc\,{\sc ii} start emerging. The Fe\,{\sc ii} lines around 5000 \AA\, also start blending. The Ca\,{\sc ii} NIR triplet gradually transformed into emission dominated. Beyond +40 d, when the light curve enters into the exponential decline, the forbidden lines [O\,{\sc i}] 6300, 6364 \AA\, and [Ca\,{\sc ii}] 7291, 7324 \AA\, start appearing. 

From the light curve, it is clear that the supernova entered into the linear decline phase beyond 50 d after the explosion, which indicates a transition from the photospheric phase to the nebular phase. The emergence of nebular lines, especially the forbidden lines of [O\,{\sc i}] 6300, 6364 \AA\, and [Ca\,{\sc ii}] 7291, 7324 \AA\, also signifies the transition to the nebular phase. The spectrum taken during +67 to +112 d is displayed in Fig.~\ref{fig_neb}. During this phase, the spectra are slowly getting dominated by the forbidden emission lines. The nebular line [O\,{\sc i}] 5577 \AA\, is also seen along with O\,{\sc i} 7774 \AA\, line, which is probably powered by an oxygen recombination cascade.
The Ca\,{\sc ii} NIR triplet is dominated by the emission component and possibly blended with the [C\,{\sc i}] 8730 \AA\, line. The spectrum obtained on +112 d still shows the presence of continuum and is not fully nebular. In Fig.~\ref{fig_comp1} (panel: C), the spectra near a month after the maximum light is compared. The $\sim$+31 d spectrum of SN~2017iro is similar to the spectrum of iPTF13bvn and SN~2008D. Also, He\,{\sc i} 6678, 7065 \AA\ lines in SN~2009jf are weaker as compared to other SNe. It is also worth mentioning that between 4000\,--5000\, \AA\, SN~2017iro and SN~2008D exhibit comparatively flat structures however, iPTF13bvn and SN2009jf have strong features in this wavelength range.

\subsection{Nebular phase}

The spectra taken during +134 to +209 d is presented in Fig.~\ref{fig_neb_late}. The spectra are dominated by the forbidden emission lines of [O\,{\sc i}] and [Ca\,{\sc ii}], narrow lines from the nearby H\,{\sc ii} region are also clearly seen. The [O\,{\sc i}] 5577 \AA\, which was fairly strong in the early nebular phase (before $\sim$\,100 d), faded after that and almost vanished in the nebular spectra. The semi-forbidden line Mg\,{\sc i}] 4571 \AA\, is found to be always weak in the spectra of SN~2017iro. The emission component of the Ca\,{\sc ii} NIR triplet, which was stronger till $\sim$\,120 d starts weakening, and by $\sim$\,176 d it merges with the almost flat continuum. The forbidden [O\,{\sc i}] and [Ca\,{\sc ii}] lines show  complex profile.

In Fig.~\ref{fig_nebular}, the nebular phase spectrum of SN~2017iro at +209 d is compared with those of other type Ib events, obtained at similar epoch. The emission features of semi-forbidden \ion{Mg}{1}] 4571 \AA, forbidden [\ion{O}{1}] 6300, 6364 \AA, and [\ion{Ca}{2}] 7291, 7324 \AA\, are prominently visible in these SNe. It is interesting to note that in SN~2017iro, the strength of [\ion{O}{1}] and [\ion{Ca}{2}] lines are similar whereas, in SN~2009jf, SN~2012au and Master OT J120451.50+265946.6 (MOT) the [\ion{O}{1}] line is stronger than the [\ion{Ca}{2}] line. In SN~2017iro, the \ion{Mg}{1}] is comparatively very weak than other SNe. The ratio of \ion{Mg}{1}] to [\ion{O}{1}] lines in SN~2017iro is estimated as $\sim$0.15 which is similar to MOT \citep[$\sim$0.3,][]{2019MNRAS.485.5438S} but lesser than iPTF13bvn \citep[$\sim$0.85,][]{2015A&A...579A..95K}.

Amongst the nebular forbidden emission lines, the \ion{Mg}{1}] line lies in the bluer and crowded region of the spectra, where generally signal is poor. Furthermore, Ca is a trace element and its distribution is not necessarily representative of the bulk of the ejecta. Hence, the [\ion{O}{1}] doublet which emerges in relatively clear region, is generally used as the tracer of the explosion geometry \citep{2005Sci...308.1284M, 2008Sci...319.1220M, 2008ApJ...687L...9M, 2009MNRAS.397..677T, 2010ApJ...709.1343M}. From the Fig.~\ref{fig_nebular}, similarity between  nebular spectrum of SN~2009jf and SN~2017iro is evident with [\ion{O}{1}] line profile being asymmetric and multi-peaked in both the objects. However, asymmetry in [\ion{O}{1}] line profile of SN~2017iro is weaker as compared to SN~2009jf with the peaked component of [\ion{O}{1}] line centered at rest wavelength. The asymmetric and multi-component [\ion{O}{1}] profile, seen in the nebular spectra, can be reproduced with a complex ejecta geometry of an aspherical explosion \citep{2005Sci...308.1284M, 2007ApJ...666.1069M}. The observed features in the asymmetric [\ion{O}{1}] line profile in SN~2017iro, could be explained with a high density core of the ejecta \citep{2009MNRAS.397..677T, 2008Sci...319.1220M}.

The gas-phase oxygen abundance in the vicinity of supernova location is estimated using the relation of \citet{2004MNRAS.348L..59P}. N2 and the O3N2 indices were computed from the last spectrum obtained on $\sim$\,209 d. The average value of oxygen abundance 12 + log(O/H) (estimated using O3N2 and N2 indices) for the supernova region is found to be 8.64 $\pm$ 0.2 which is remarkably similar to iPTF13bvn \citep[8.63,][]{2015A&A...579A..95K}. In a recent study under the CALIFA \citep{2012A&A...538A...8S} survey program, \citet{2016A&A...591A..48G} used wide-field integral field spectroscopy to compute the global metallicity of NGC~5480 and metallicity at the location of SN~1988L. The estimated values are  8.58 and 8.55 for the local and global, respectively. This shows that our estimate of the oxygen abundance is in fairly good agreement with the previous study, and is close to the solar value of 12 + log (O/H) = 8.69 $\pm$ 0.05 \citep{2009ARA&A..47..481A}. Further, the computed metallicity close to the SN location is similar to the estimates of \citet[][8.49\,$\pm$\,0.19]{2011ApJ...731L...4M} and \citet[][8.48\,$\pm$\,0.16]{2012ApJ...758..132S}, made at the location of other SNe Ib.

\subsection{Line velocities}\label{vel}

The velocities of lines  Fe\,{\sc ii} 5169 \AA, He\,{\sc i} 5876 \AA, He\,{\sc i} 6678 \AA\ and Si\,{\sc ii} 6355 \AA\ were estimated by fitting  Gaussian profile to the respective absorption trough after correcting the spectra for the redshift of host galaxy of SN~2017iro. The estimated line velocities of SN~2017iro along with SNe~2007Y, 2008D, 2009jf, 2012au, and iPTF13bvn collected from the literature are plotted in Fig.~\ref{fig_c2}. In case of SN~2017iro, the He\,{\sc i} 5876 \AA\, velocity is $\sim$11500 km s$^{-1}$ in the beginning and declined to $\sim$9000 km s$^{-1}$ near maximum. In the post-maximum phase, it evolves slowly and becomes flat ($\sim$7000 km s$^{-1}$) after +70 d. All the SNe (except SN~2007Y), used in the comparison, show a steep decline in the He\,{\sc i} 5876 \AA\ line velocity during the pre-maximum phase. The post-maximum evolution of He\,{\sc i} shows a modest increase of $\sim$900 $\rm km\ s^{-1}$ giving rise to a bump like feature in the velocity evolution. A similar feature can also be seen in SN~2009jf, iPTF13bvn, and SN~2008D during maximum to $\sim$+20 d. The origin of this feature needs to be explored. The evolution of He\,{\sc i} 5876 \AA\, line velocity beyond $\sim$+25 d, is similar in SN~2017iro, iPTF13bvn, and SN~2009jf but lower than SN~2012au. In homologous expanding SNe ejecta, it is difficult to accurately determine the photospheric velocity from the observed spectral features as none measurable feature of the spectra is connected directly to photospheric velocity \citep{2012MNRAS.419.2783T}.However, the prominent absorption features produced due to Fe\,{\sc ii} 5169 \AA\, can be used as a good tracer of SN photospheric velocity as it is a less optically thick line compared to several others lines \citep{2005Dessart}. The Fe\,{\sc ii} line velocity of SN~2017iro is compared in the middle panel of Fig.~\ref{fig_c2}. The ejecta velocity measured using Fe\,{\sc ii} line of all the SNe shows a steep decline till $\sim$+25 d and flattens thereafter. The Fe\,{\sc ii} line velocity of SN~2017iro near-maximum is comparable to SN~2007Y, SN~2009jf, and iPTF13bvn, but slower than SN~2008D and SN~2012au. In the later phase ($>$+25 d), the Fe\,{\sc ii} line velocity of SN~2017iro is in between SN~2009jf and SN~2007Y. The Fe\,{\sc ii} velocity around maximum light ($\sim$9000 $\pm$ 800 km s$^{-1}$) is used to estimate the explosion energy of SN~2017iro (see Section~\ref{exp_pa}).

\section{Discussion}\label{diss}

The observed fraction of Type Ib SNe is low, they are rare objects \citep{2011-Li, 2017Graur-1, 2017Graur-2, 2019Shivvers}. There are only a handful events for which studies covering both photospheric and nebular phases are available. A detailed investigation of SN~2017iro, hence is a good addition to this sample as it fills the gap between fast declining (e.g. SN~2007Y, iPTF13bvn) and slow declining (e.g. SN~2009jf) objects (see Section~\ref{lc_fit}). Here, we discuss the nature of SN~2017iro based on some significant light curve and spectral features, constrain its progenitor mass, and also estimate the explosion parameters.

\subsection{Late phase light curve heterogeneity}\label{late_lc}

The light curve of SE-SNe follows a linear decline with the onset of the nebular phase (usually $>$\,100 days after the explosion). This is a consequence of energy injection from the radioactive decay of $^{56}$Co $\rightarrow$ $^{56}$Fe. The efficiency of $\gamma$-ray trapping within the ejecta affects the decay rate. In general, it is found that the late phase light curve of SE-SNe declines faster than the expected decay rate of 0.98 mag (100 d)$^{-1}$ from $^{56}$Co $\rightarrow$ $^{56}$Fe transition. This is primarily due to the lower opacity of the ejecta for the $\gamma$-rays (i.e., incomplete trapping). Therefore, the light curve decay rate during the nebular phase can be used to infer the ratio of explosion energy, and mass of the ejecta. The late phase (beyond 140 d) decline rate of SN~2017iro is estimated in different bands and are listed in Table~\ref{tab_lc_p}. The light curves are found to decline significantly faster than 0.98 mag (100 d)$^{-1}$.

An investigation of the late time light curve of SE-SNe by \citet*{2015MNRAS.450.1295W} revealed the diverse nature of their evolution, indicating dispersion in the ejecta masses and kinetic energies. Our comparison of the late-time decay rate of well studied Type Ib SNe (cf. Fig.~\ref{fig_lc}: right panel) indicates that there exist two groups of events. SN~1999dn and SN~2012au fall under the first category in which the late phase decline rate is significantly slower than the $^{56}$Co $\rightarrow$ $^{56}$Fe decay rate. The slower light curve decline rate indicates the need for an additional source of energy which could be interaction of the ejecta with the circumstellar material (CSM), a magnetar or presence of light echo. \citet{2015MNRAS.450.1295W} suggested the presence of CSM in SN~1999dn. The presence of smoothly distributed CSM around SN~2012au was inferred from its detection in radio wavebands \citep{2014ApJ...797....2K}. However, very late nebular spectrum of SN 2012au, obtained $\sim$\,6.2 years after the explosion, indicated magnetar/pulsar wind nebulae as the most probable energy source \citep{Milisavljevic-2018}. In the second group of SE-SNe, the light curve decline rate is faster than the expected decline rate of $^{56}$Co\,--\,$^{56}$Fe. The faster decline rate could be due to incomplete trapping of $\gamma$-rays produced via $^{56}$Co\,--\,$^{56}$Fe decay, in the SN ejecta. The late-time decay rate of SN~2017iro indicates it falls in this category and is similar to that of SN~2009jf, SN~2007Y and iPTF13bvn. It is worth mentioning that during the late nebular phase, an increasing fraction of photons emitted at IR wavelength may also steepen the decline rate in optical bands.


\subsection{$^{56}$Ni mixing}\label{He_diss}

The lines of He\,{\sc i}, the identifying features of Type Ib SNe, are seen at a velocity of $\sim$\,11500 km s$^{-1}$ in the first spectrum ($\sim$--7 d) of SN~2017iro (cf. Fig.~\ref{fig_c2}). The origin of these lines require non-thermal excitation and ionization \citep{1991ApJ...383..308L,1991ApJ...373..604S}. The accelerated electrons responsible for the non-thermal excitation are regulated by the $\gamma$-rays produced in the radioactive decay of newly synthesized $^{56}$Ni \citep{1987ApJ...317..355H}. It is shown that the mere presence of an appreciable amount of helium in the progenitor would not lead to the presence of helium lines in the spectra \citep{2012MNRAS.424.2139D}. To excite He\,{\sc i} at high velocities, $\gamma$-rays need to be in the proximity of the helium layer, either due to the leakage of $\gamma$-rays through the inner ejecta or by substantial mixing of $^{56}$Ni up to the outer ejecta. 

Mixing may arise due to various reasons such as large-scale asymmetric explosion jets and asymmetry of the propagating shock etc. \citep[see,][and references therein]{Maund2009, Couch2011, 2012MNRAS.424.2139D}. The effect of $^{56}$Ni mixing on the light curves and spectra of SE-SNe have been investigated in detail by several authors. Theoretical studies \citep{1990ApJ...361L..23S, 1991ApJ...368L..27H} argued that $^{56}$Ni mixing within the helium layer might depend on the progenitor composition. The degree of $^{56}$Ni mixing may also significantly influence the overall light curve shape (including the post-maximum phase) and the efficient formation of He\,{\sc i} spectral lines \citep{1988ApJ...333..754E, 1997ASIC..486..821W, 1990ApJ...361L..23S, 2014MNRAS.438.2924C, 2015MNRAS.453.2189D, 2016MNRAS.458.1618D, 2018AA...609A.136T}.

\begin{figure}
\centering
\includegraphics[width=\columnwidth]{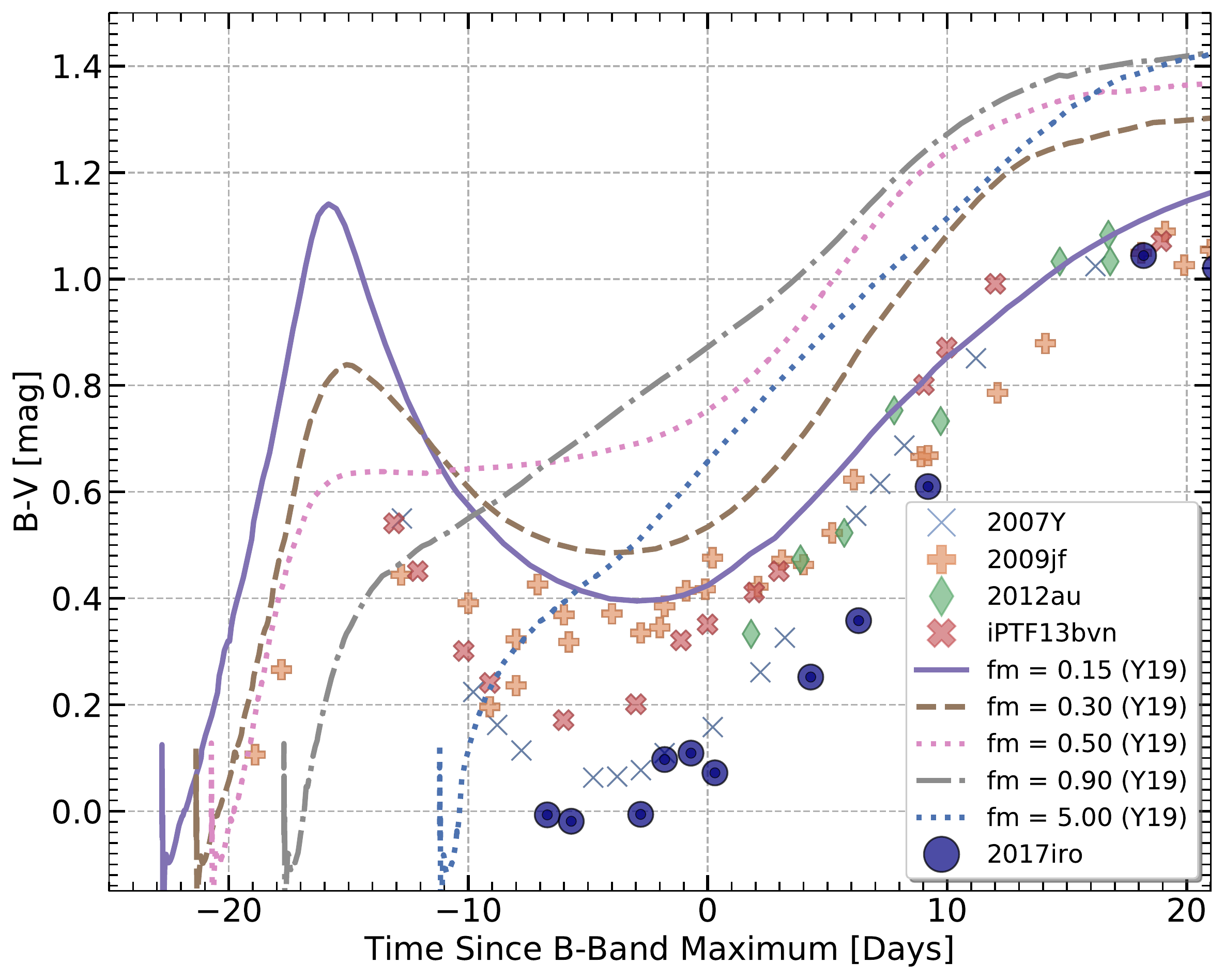}
\caption{The $B-V$ colour evolution of SN~2017iro compared with the synthetic colour curves of SN Ib models by \citet{2019ApJ...872..174Y}. The five bold curves indicate varying degrees of $^{56}$Ni mixing (f$_{m}$) as shown in the legend (Y19). An increasing value of f$_{m}$ implying a stronger $^{56}$Ni mixing \citep[for more details, see][]{2019ApJ...872..174Y}.}
\label{fig_yoon}
\end{figure}

The early time colour evolution of Type Ib/c SNe can also be used as a diagnostic for the $^{56}$Ni distribution in the ejecta. It is shown that in the case of strong mixing, the colour curves of Type Ib/c SNe monotonically redden during the photospheric phase, because of continuous and progressive effects of $^{56}$Ni heating. Whereas, in the case of weak to moderate mixing, the colour evolution does not show monotonic behaviour due to delayed heating from $^{56}$Ni \citep[][and references therein]{2019ApJ...872..174Y}. The early phase $B-V$ colours of different Ib events along with $^{56}$Ni mixing models \citep{2019ApJ...872..174Y} are plotted in Fig.~\ref{fig_yoon}.
The colour evolution of SN~2007Y, SN~2009jf, and iPTF13bvn follow a non-monotonic trend which is interpreted in the framework of weak to moderate mixing of $^{56}$Ni in the ejecta. However, the non-availability of sufficiently early phase data points for SN~2017iro prevents us from arriving at any firm conclusion on the role of mixing.

\subsection{Physical parameters of explosion}\label{exp_pa}

The peak of the light curve and its shape in SE-SNe are mainly regulated by the synthesized radioactive $^{56}$Ni, kinetic energy ($E_{\rm k}$) and ejecta mass ($M_{\rm ej}$). Estimation of these parameters are crucial in understanding the explosion properties. The explosion parameters can be derived employing either detailed hydrodynamical modelling of both the light-curve and the spectra or using semi-analytical models. In the present study, several semi-analytical formulations were used to fit the quasi-bolometric light curve (see Section~\ref{bol}) of SN~2017iro.

\subsubsection{Estimation of $^{56}$Ni mass, ejecta mass and kinetic energy}

The optically thick phase of the light curve (photospheric phase), was fitted by the formulation originally proposed by \citet{1982ApJ...253..785A} and later updated by \citet{2008MNRAS.383.1485V}. Major assumptions in this analytical approach comprise a small radius of the progenitor, spherical symmetry and homologous expansion of the ejecta, constant opacity ($\kappa_{\rm opt}$) and a centrally located and unmixed $^{56}$Ni \citep[see][]{1982ApJ...253..785A, 2008MNRAS.383.1485V, 2013MNRAS.434.1098C}. The parameters $M_{\rm Ni}$ (nickel mass) and $\tau_{\rm m}$ (diffusion time-scale) were kept as free variables while obtaining the fit. Considering a uniform density medium, the ejecta kinetic energy $E_{\rm k}$ and $\tau_{\rm m}$ are, given by
\begin{equation}\label{eq_1}
\tau_{\rm m} = \sqrt{2} \left( \frac{\kappa_{\rm opt}}{\beta c} \right)^{1/2} \left( \frac{M_{\rm ej}}{v_{\rm ph}} \right)^{1/2},
\end{equation}

\begin{equation}\label{eq_2}
E_{\rm k} \approx \frac{3}{5}\frac{M_{\rm ej}v^2_{\rm ph}}{2},
\end{equation}

where $\beta \approx 13.8$ is a constant of integration \citep{1982ApJ...253..785A} and {\it c} is the speed of light. The optical opacity $\kappa_{\rm opt}$ is adopted as 0.07 cm$^2$ g$^{-1}$ \citep[e.g.][]{2000AstL...26..797C, 2016ApJ...818...79T,2018AA...609A.136T}. The quasi-bolometric light-curve of SN~2017iro until 30 d from the explosion epoch was fitted using least-square optimization. The best fit was obtained for a $M_{\rm Ni}$ = 0.09 $\pm$ 0.04 M$_{\odot}$ and $\tau_{\rm m}$ = 9.95 $\pm$ 0.44 d. The derived $^{56}$Ni mass was further estimated using the bolometric rise time of 16.5 d and a peak bolometric luminosity log$_{10}L$\,=\,42.39 erg s$^{-1}$. The relation given by \citet{Stritzinger-2005} gives a value of 0.10 $\pm$ 0.01 M$_{\odot}$. Using the measured Fe\,{\sc ii} line velocity 9000 km s$^{-1}$ near the bolometric maximum (see Section~\ref{pre_max_spc}), the $M_{\rm ej}$ and $E_{\rm k}$ are estimated as 1.39 M$_{\odot}$ and 0.75\,$\times$\,10$^{51}$ erg, respectively.

\begin{figure}
\centering
\includegraphics[width=\columnwidth]{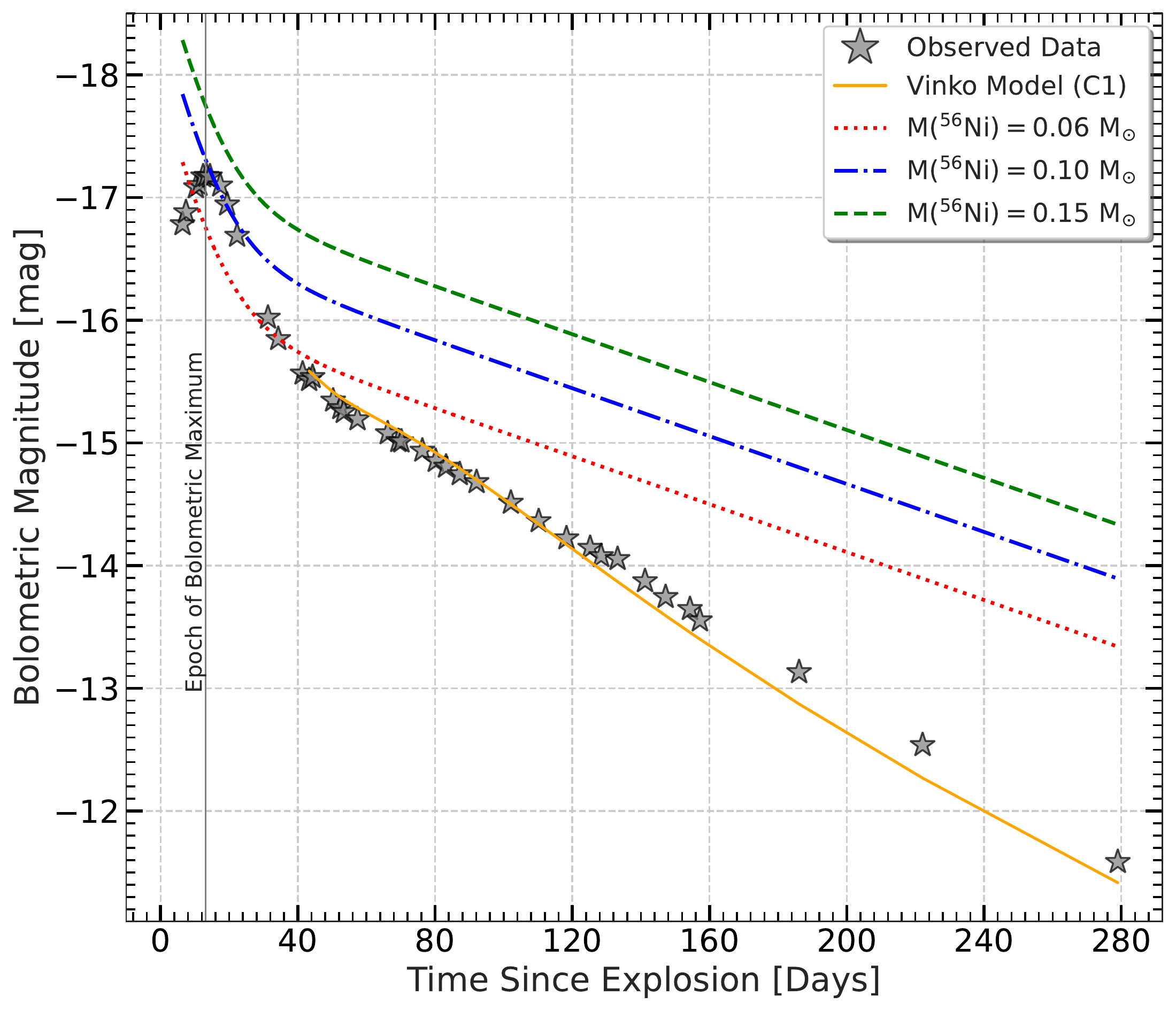}
\caption{The quasi-bolometric light curve of SN~2017iro fitted with analytical model C1 of \citet{2004A&A...427..453V}. Three curves (dotted, dashed, and dash-dotted) representing the rate of energy production for different masses of $^{56}$Ni synthesized during the explosion, based on the analytical formulation by \citet{1994ApJS...92..527N} have also been shown.}
\label{bol_expl}
\end{figure}

While investigating the physical parameters of SN~2002ap, \citet{2004A&A...427..453V} proposed analytic models for post-maximum phase with assumptions as described in the beginning of this sub-section. In brief, these models are based on the ejecta density configuration characterized by $x_{0}$ (fractional radius of the core) and $n$ (power-law exponent) parameters. We used their model C1 which represents a `core-shell' structure with a fixed density core of fractional radius, $x_{0}$\,=\,0.15 and a fixed value of $\kappa_{\gamma}$ = 0.027 cm$^2$ g$^{-1}$ for grey atmospheres \citep{1984ApJ...280..282S}. The best-fit model is shown in Fig.~\ref{bol_expl} with an $M_{\rm ej}$ = 4.3 M$_{\odot}$, $M_{\rm Ni}$ = 0.05 M$_{\odot}$, $E_{\rm k}$ = 1.85 foe and $v_{\rm max}$ = 8,500 $\rm km s^{-1}$. 

Energy production rates of $^{56}$Ni $\rightarrow$ $^{56}$Co $\rightarrow$ $^{56}$Fe decay chain computed by \citet{1994ApJS...92..527N} were further used to examine the mass of synthesized $^{56}$Ni during the explosion. The energy production curves for three different values (i.e. 0.06, 0.10, and 0.15 M$_{\odot}$) of the mass of $^{56}$Ni were over-plotted in Fig.~\ref{bol_expl}. The maximum in the quasi-bolometric light curve describes roughly the transition between the emission deficit due to the large optical depth (larger diffusion timescale) and an excess emission due to the stored radiation post-maximum. The early post-maximum decline can be approximated by the instantaneous energy production rate of $^{56}$Ni $\rightarrow$ $^{56}$Co decay. Hence, we infer that a 0.10 M$_{\odot}$ of $\rm ^{56}Ni$ was synthesised in the explosion due to its best-match to the initial post-maximum decline of the quasi-bolometric light-curve of SN~2017iro. The large discrepancy between the energy production rate and the quasi-bolometric luminosity during the late phase is arising due to the escape of gamma-rays.

\begin{figure}
\centering
\includegraphics[width=\columnwidth]{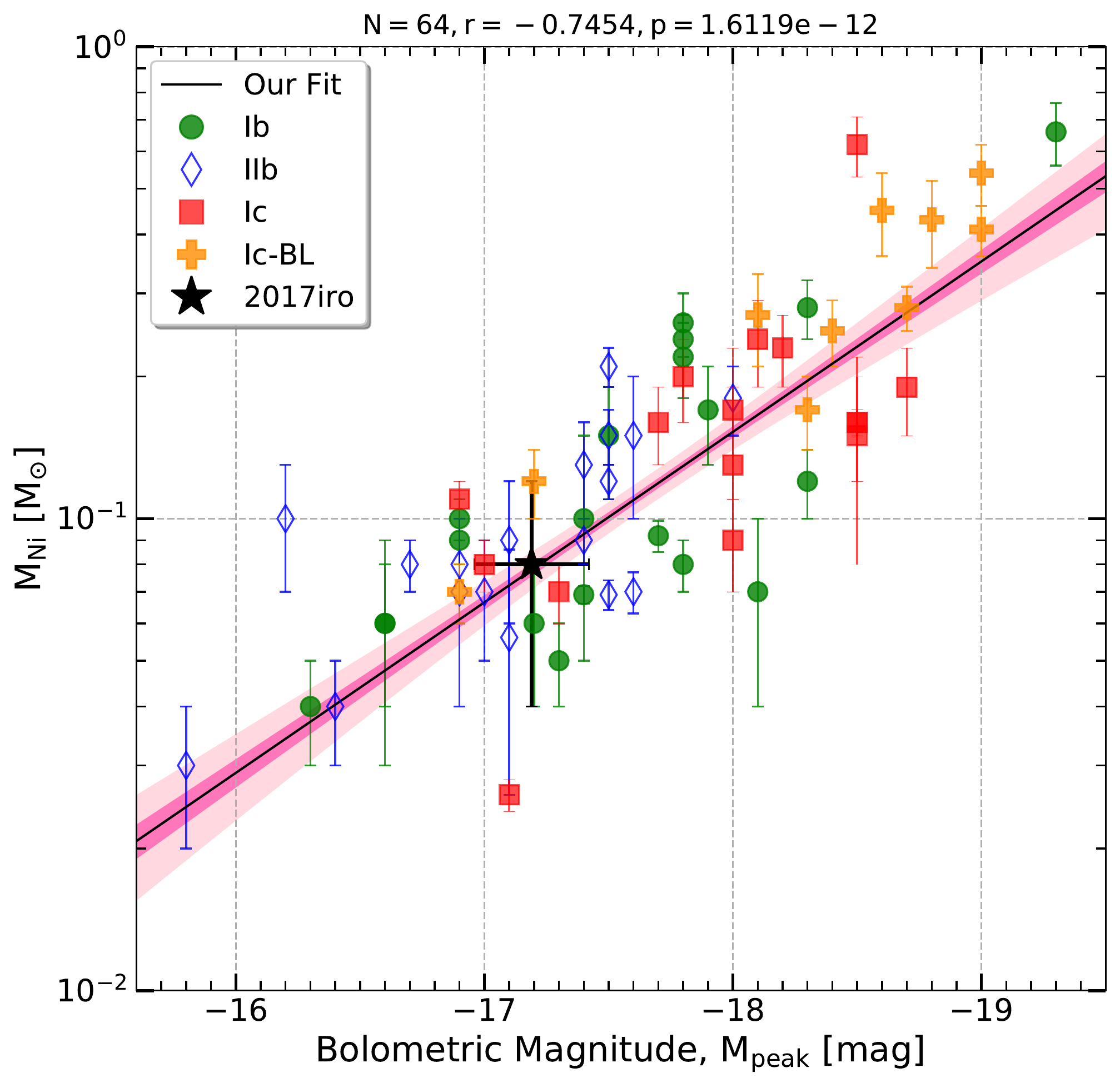}
\caption{Correlation between M$_{bol}$ and $^{56}$Ni mass of 47 SE-SNe (including the sample from \citet{2016MNRAS.457..328L} and 9 additional events, see Section~\ref{exp_pa}). Type Ib, Ic, Ic-BL and IIb are shown with different symbols. The bold continuous line is best fit. The dark--pink and light--pink shaded regions are indicative of 1$\sigma$ and 3$\sigma$ confidence intervals of the fit, respectively.}
\label{ni_fit}
\end{figure}

Based upon a study of 38 SE-SNe, \citet{2016MNRAS.457..328L} proposed a fitting relation to estimate $M_{\rm Ni}$ (see their Eq. 4). To populate this data sample, we collected 9 additional SE-SNe (5 IIb, 2 BL-Ic, 1 Ic-transitional, and 1 Ib events) with well-sampled light curves from the literature (see Table~\ref{add_sn}). Also, to estimate M$_{bol}$, the data available only in the optical wavelength was used. In Fig.~\ref{ni_fit}, a correlation between M$_{bol}$ and $^{56}$Ni mass is shown for all the 47 events. 
The plot also indicates that among the population of SE-SNe that the Type IIb events produce a lesser amount of $^{56}$Ni. However, there appears to be no significant difference in $^{56}$Ni production between Type Ib and Ic SNe, whereas Type Ic-BL SNe, in general, are brighter with higher $^{56}$Ni mass. This is in line with other recent studies \citep{2019A&A...628A...7A, 2019MNRAS.485.1559P}. With a $M_{bol_{peak}}$ of --17.19 $\pm$ 0.22 mag (cf. Table~\ref{tab_lc_p}) and $^{56}$Ni mass of 0.08 M$_{\odot}$ (average value estimated from above methods), SN~2017iro (shown with a star symbol in the Fig.~\ref{ni_fit}) lies near the expected correlation for SE-SNe. It is to be noted here that the explosion parameters inferred using empirical relations are approximate, and detailed hydrodynamical modelling \citep[e.g.][]{2018AA...609A.136T} would be required for better estimates.

\begin{table}
\scriptsize
\centering
\caption{In addition to \citet{2016MNRAS.457..328L} following events are used to estimate correlation between M$_{bol}$ and $^{56}$Ni mass (see Section~\ref{exp_pa}).}\label{add_sn}
\begin{tabular}{lllcc}
\hline
Name        & Type  &   M$_{bol}$    & $^{56}$Ni mass    & References  \\
            &       &   (mag)        &  (M$_{\odot}$)    &             \\
\hline
SN~2011ei   & IIb   &  --\,15.8      & 0.03 $\pm$ 0.01   &  1          \\
SN~2011fu   & IIb   &  --\,17.5      & 0.21 $\pm$ 0.02   &  2          \\
SN~2012ap   & Ic-BL &  --\,17.2      & 0.12 $\pm$ 0.02   &  3          \\
SN~2012au   & Ib    &  --\,17.8      & 0.26 $\pm$ 0.04   &  4, 5, 6    \\
SN~2013df   & IIb   &  --\,16.2      & 0.10 $\pm$ 0.03   &  7, 8       \\
SN~2014ad   & Ic-BL &  --\,18.1      & 0.27 $\pm$ 0.06   &  9          \\
SN~2015as   & IIb   &  --\,16.7      & 0.08 $\pm$ 0.01   &  10         \\
SN~2016coi  & Ic/Ic-BL$^\ast$&  --\,16.9      & 0.11 $\pm$ 0.01   & 11 \\
SN~2016gkg  & IIb   &  --\,17.4      & 0.09 $\pm$ 0.01   &  12         \\
\hline
\end{tabular}\\
References:
$^{1}$\,\citep{2013ApJ...767...71M},
$^{2}$\,\citep{2013-Kumar},
$^{3}$\,\citep{2015ApJ...799...51M},
$^{4}$\,\citep{2013ApJ...770L..38M},
$^{5}$\,\citep{Pandey_12au},
$^{6}$\,\citep{2013ApJ...772L..17T},
$^{7}$\,\citep{2014MNRAS.445.1647M},
$^{8}$\,\citep{2016MNRAS.460.1500S},
$^{9}$\,\citep{2018MNRAS.475.2591S},
$^{10}$\,\citep{2018MNRAS.476.3611G},
$^{11}$\,\citep{2018MNRAS.473.3776K},
$^{12}$\,\citep{2017ApJ...836L..12T}.\\
$^\ast$ Ic/Ic-BL transitional event.
\end{table}

\begin{table}
\scriptsize
\centering
\caption{Oxygen mass of Type Ib SNe.}
\begin{tabular}{lcc} \hline
Name       & Oxygen mass (M$_\odot$) & References \\ \hline
SN~1996N   & 0.11\,--\,0.21          & 1          \\
SN~2007Y   & 0.20                    & 2          \\
SN~2009jf  & 1.34                    & 3          \\
iPTF13bvn  & 0.33                    & 4          \\
MOT$^{*}$  & 0.90                    & 5          \\
SN~2017iro & 0.35                    & This study \\
\hline
\end{tabular}\\
$^{*}$ Master OT J120451.50+265946.6 \\
References:
$^{1}$\,\citep{1998A&A...337..207S},
$^{2}$\,\citep{2009ApJ...696..713S},
$^{3}$\,\citep{2011MNRAS.413.2583S},
$^{4}$\,\citep{2015A&A...579A..95K},
$^{5}$\,\citep{2019MNRAS.485.5438S}
\label{sn_Omass}
\end{table}

\subsubsection{Oxygen mass} \label{sec:oxygen}

The [\ion{O}{1}] line flux in the nebular spectra of SE-SNe can be used to estimate the neutral oxygen mass that is required to produce the emission. In the high electron density ($n_{e}\,\ge 10^{-6}$ cm$^{-3}$) region, the minimum O mass can be estimated by the relation $M_{O} = 10^{8} \times D^{2} \times F([\rm{OI}]) \times exp{(2.28/T_{4})}$, provided by \citet{1986ApJ...310L..35U}. Here, mass of neutral oxygen is represented as $M_{O}$ (in $\rm M_{\odot}$), $D$ is the SN distance (in Mpc), $F$([\ion{O}{1}]) is the observed absolute flux of [\ion{O}{1}] line (in erg s$^{-1}$ cm$^{-2}$) and $T_4$ is the temperature associated with oxygen-emitting region in units of 10$^{4}$ K. The [\ion{O}{1}] 5577\,/\,(6300, 6364)\, line ratio can be used as a proxy for estimating the temperature. In our nebular phase spectra (cf. Fig.~\ref{fig_neb_late}), the [\ion{O}{1}] 5577 \AA\, line is not detected, allowing us to set $\sim$\,0.1 as the upper limit for the line ratio. At this limit, the emitting region should either be at a relatively low temperature ($T_{4} \le 0.4$) for the high density limit or be at a low electron density ($n_{e} \le$ 5 $\times$ 10$^{6}$ cm$^{-3}$) if T$_{4}$ = 1 \citep{2007ApJ...666.1069M}.

Previous studies have shown that during the nebular phase, the temperature of the line emitting region lies in the range 3400\,--\,4200 K \citep{1989AJ.....98..577S, 2004A&A...426..963E, 2011AcA....61..179E}. Assuming that the high-density regime is valid for SN~2017iro, the temperature of the line emitting region can be taken to be $T_{4}$\,=\,0.4. Using the [\ion{O}{1}] flux (in the +209 d spectrum) of 1.12 $\times$ 10$^{-14}$ erg s$^{-1}$ cm$^{-2}$, D\,=\,30.8 Mpc and $T_{4}$\,=\,0.4, we calculated the minimum mass of neutral oxygen to be 0.35 M$_\odot$, which is similar to the mass estimated for supernova iPTF13bvn \citep{2015A&A...579A..95K}. It is to be noted that for different densities, the temperature of line emitting regions may vary \citep{1989AJ.....98..577S, 1991ApJ...372..531L, 1998A&A...337..207S, 2004A&A...426..963E, 2007ApJ...666.1069M}. An increase in temperature from 4000 K to 4200 K lowers the oxygen mass by a factor of $\sim$\,0.8. Similarly, a decrease in temperature from 4000 K to 3400 K results in higher oxygen mass by a factor of $\sim$\,3. Hence, depending on the assumed temperature and density, the range of oxygen mass spans 0.28 -- 1.05 M$_{\odot}$. The clumpy nature of the ejecta and the presence of non-optically thin material (which is not accounted here) may raise this value \citep{2007ApJ...658L...5M, 2015A&A...579A..95K}. In Table~\ref{sn_Omass}, the neutral oxygen mass estimates are listed for a few well-studied objects showing a broad range of estimated masses. We admit that the mass of the oxygen estimated here is a lower limit of the total oxygen present in the ejecta of SN~2017iro.

\begin{figure}
\centering
\includegraphics[width=\columnwidth]{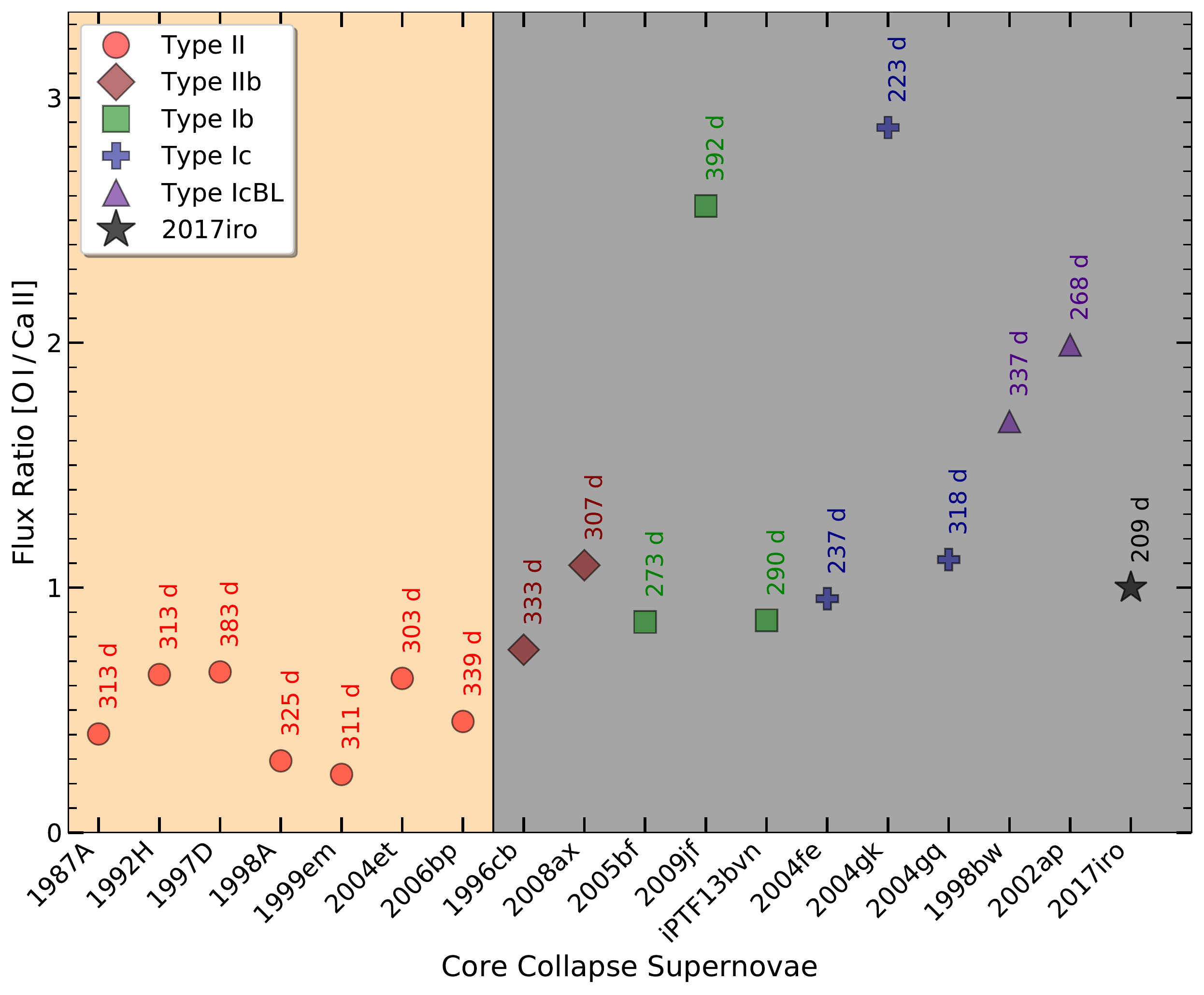}
\caption{[\ion{O}{1}] 6300, 6364\,/\,[\ion{Ca}{2}] 7291, 7324 \AA\ line ratio of CCSNe from \citet{2015A&A...579A..95K}. The [\ion{O}{1}]/[\ion{Ca}{2}]\, ratio for SN~2017iro obtained from the nebular spectra (between +173 to +209 d) has been shown with a star symbol. The phases (relative to the time of maximum light) are shown next to the each data point. The left (orange) and right (grey) regions belong to Type II and SE-SNe, respectively. The reference for spectrum are the following: SNe~1987A \citep{1995Pun}, 1992H \citep{1996Clocchiatti}, 1997D \citep{2001Benetti}, 1998A \citep{2005Pastorello}, 1998bw \citep{2001Patat}, 1999em \citep{2002Leonard}, 2004et \citep{2006Sahu}, 2004gq \citep{2008SMaeda}, 2005bf \citep{2007Maeda}, 2006bp \citep{2007ApJ...666.1093Quimby}, 1996cb, 2002ap, 2004fe, 2004gk, 2008ax, 2009jf \citep{2008Modjaz,2014Modjaz}.}
\label{fig_oicaii}
\end{figure}

\subsection{Nature of the progenitor star}

Hydrodynamic explosion models suggest that Mg and O should have similar spatial distribution within the SN ejecta of SE-SNe \citep{2006ApJ...645.1331M}. \citet{2009MNRAS.397..677T} analyzed a large number of nebular spectra of SE-SNe and inferred that in a majority of cases, the line profiles of semi-forbidden \ion{Mg}{1}] 4571 \AA\, and forbidden oxygen [\ion{O}{1}] 6300, 6364 \AA\ are similar, confirming the similar distribution of Mg and O in the ejecta. The strength of \ion{Mg}{1}] 4571 \AA\ line grows over time in comparison with [\ion{Ca}{2}] 7291, 7324 \AA\, and [\ion{O}{1}] because the deeper Mg-O layer of the progenitor core becomes visible with time \citep{2003-Foley,2015A&A...579A..95K}. SN~2017iro shows a feeble presence of \ion{Mg}{1}] 4571 \AA\, possibly due to less stripping of its progenitor. 

The progenitor mass of a CCSN can be constrained by estimating the [\ion{O}{1}] 6300, 6364\,/\,[\ion{Ca}{2}] 7291, 7324 \AA\ emission line ratio in the nebular phase \citep{2015A&A...579A..95K}. In contrast to Ca, the explosively synthesized O is sensitive to the main sequence mass of the progenitor, and therefore, a larger value implies a larger core mass \citep[however, mixing can also play a substantial role, see][]{1989ApJ...343..323F}. \citet{2015A&A...579A..95K} compared the [\ion{O}{1}]\,/\,[\ion{Ca}{2}] line ratio of SE-SNe and Type II SNe. They found that the line ratio of Type Ib/c SNe exhibits considerable dispersion. The observed spread in the line ratio can not be explained by the temporal evolution of the line flux. Instead, it could arise from two different progenitor channels for SE-SNe: a single massive star or a less massive star in a binary system. The [\ion{O}{1}]\,/\,[\ion{Ca}{2}] ratio in the spectra of SN~2017iro obtained between 173 to 209 d stays almost constant around 1. In Fig.~\ref{fig_oicaii}, the [\ion{O}{1}]\,/\,[\ion{Ca}{2}] ratio of CCSNe (from \citet{2015A&A...579A..95K}) is plotted where SN~2017iro is shown with a star symbol. It indicates that the progenitor of SN~2017iro is comparatively a less massive star in a binary system. A less massive progenitor in a binary system has been proposed in several Type Ib supernovae such as SN~2007Y \citep{2009ApJ...696..713S}, iPTF13bvn \citep{2014AJ....148...68B, 2015A&A...579A..95K, 2015MNRAS.446.2689E}. This is in agreement with the fact that binary interaction is a favoured scenario in a majority ($\ge$70\%) of massive stars \citep{1992ApJ...391..246P, 2012Sci...337..444S}. 

The nucleosynthesis calculations for zero-age sequence progenitor masses of 13, 15, 20, and 25 M$_{\odot}$ was performed by \citet*{1996ApJ...460..408T}, and the corresponding oxygen mass was estimated as 0.22, 0.43, 1.48, and 3.00 M$_{\odot}$, respectively. \citet{2003limongi} also computed explosive yields of massive stars in the mass range 13\,--\,35 $\rm M_{\odot}$ with an initial solar composition using the FRANEC \citep{1998cheiffi} code. Their estimated oxygen mass is $\sim$0.3 and $\sim$0.5 M$_{\odot}$ for 13 and 15 M$_{\odot}$ progenitors masses, respectively. Both the nucleosynthesis calculations indicate that the estimated oxygen mass of 0.35 M$_\odot$ could be produced by a progenitor star with zero-age main sequence mass in the range $\sim$13--15 M$_\odot$. It is further supported by the findings of \citet{Fang2019}, where the upper limit of the Type IIb/Ib SNe progenitors is proposed as $\sim$17 M$_\odot$ for a progenitor in a binary system \citep[see, also][]{2018AA...609A.136T, 2019MNRAS.485.1559P}. It is worth mentioning that the estimated oxygen mass for SN~2017iro (Section~\ref{sec:oxygen}) is only a lower limit of the total oxygen mass. With the increased uncertainty range (0.28 -- 1.05 M$_{\odot}$), the range of possible progenitor masses increases substantially, especially towards the higher progenitor masses ($\sim$20 M$_{\odot}$). However, the lower value of [\ion{O}{1}]\,/\,[\ion{Ca}{2}] ratio indicates towards a lower zero-age main sequence mass progenitor of SN~2017iro.

The pre-explosion images of NGC~5480 available in the HST archive (observed on 2017 January 17, Proposal ID: 14840, PI: Andrea Bellini), were examined to identify the possible progenitor candidate of SN~2017iro. The image was obtained with ACS/WFC in filter F606W. A diffuse source with an extent of 5 pixels was detected in the pre-explosion images at the supernova location. At the adopted distance of the host galaxy, the extent corresponds to $\sim$\,35 parsecs. This extended source could possibly be a star cluster hosting the progenitor of SN~2017iro.


\section{Summary}\label{sum}

We have presented optical photometric (40 epochs) and spectroscopic (34 epochs) follow-up of the Type Ib SN~2017iro. The light curve evolution in $UBVRI$ follows the usual trend of SE-SNe, i.e., blue pass-bands peak before the red pass-bands indicating that the SN photosphere is rapidly cooling. The $\rm \Delta m_{15}$ parameter (for $BVRI$-bands) are comparable to those estimated for a large SE-SNe sample \citep{2011ApJ...741...97D, 2018AA...609A.136T} and in particular to SN~2012au. Further, the colours of SN~2017iro evolve similar to Type Ib events. The absolute $V$-band and bolometric (log$_{10}L$) luminosities are --17.76 mag and 42.39 erg s$^{-1}$, respectively, indicative of a moderately luminous SN. The explosion parameters were computed by applying different analytical models to the quasi-bolometric light curve which yield a $^{56}$Ni mass of $\sim$0.05\,--\,0.10 M$_{\odot}$, $M_{\rm ej}$ of $\sim$1.4\,--\,4.3 M$_{\odot}$ and $E_{\rm k}$ $\sim$0.8\,--\,1.9 $\times$ 10$^{51}$ erg.

There seem to be two groups of Type Ib SNe in terms of the light curve tail decline rates, suggesting a heterogeneity \citep{2015MNRAS.450.1295W} in the late time light curve evolution. However, lack of an adequate amount of data during the late phase of these events could be a possible reason for such inference. Hence, late phase monitoring of a larger sample of such events may provide a clearer insight to this finding. The late phase light curves of SN~2017iro decayed with moderate steepness, indicating incomplete $\gamma$-ray trapping by the SN ejecta. A low to moderate level of $^{56}$Ni mixing is supported by the non-monotonic evolution of colour curves \citep{2012MNRAS.424.2139D, 2019ApJ...872..174Y} in the case of SN~2017iro. However, we can not draw a firm conclusion due to unavailability of sufficient early phase data of this event.

In the first optical spectrum ($-7$ d) of SN~2017iro, the He\,{\sc i} 5876 \AA\ feature is clearly identified. Other He\,{\sc i} lines such as 4471, 6678, and 7065 \AA\, are comparatively weak in the beginning, but they never became as strong as He\,{\sc i} 5876 \AA. The ejecta velocity estimated using the Fe\,{\sc ii} 5169 \AA\, line is $\sim$\,9000 km s$^{-1}$ near $B$-band maximum which became almost constant to $\sim$\,5500 km s$^{-1}$ beyond +25 d. The Mg\,{\sc i}] line is very weak during the nebular phase in SN~2017iro. The general spectroscopic behaviour of SN~2017iro is similar to SN~2009jf and iPTF13bvn. The metallicity near SN~2017iro location (12 + log (O/H) = 8.64) is similar to the solar value. A neutral oxygen mass of $\sim$\,0.35 M$_\odot$ was estimated in the ejecta. The [O\,{\sc i}]\,/\,[Ca\,{\sc ii}] line ratio observed in SN~2017iro indicates the progenitor had a main sequence mass of $\sim$13\,--\,15 M$_\odot$. The inferred neutral oxygen mass and the progenitor main sequence mass for SN ~2017iro are very similar to the type Ib supernova iPTF13bvn that occurred in a binary system.

\section*{Acknowledgments}
We thank the referee for critical review and constructive suggestions that helped to improve the paper. We also thank Sung-Chul Yoon and Hanindyo Kuncarayakti for sharing the data. BK, DKS, and GCA acknowledge BRICS grant DST/IMRCD/BRICS/PilotCall1/MuMeSTU/2017(G) for the present work. DKS and GCA also acknowledge DST/JSPS grant, DST/INT/JSPS/P/281/2018. We are grateful to the observers at HCT who provided their valuable time to monitor this event and also thank the staff of IAO, Hanle, and CREST, Hosakote, that made these observations possible. The Weizmann interactive supernova data repository (WISeREP) - \url{http://wiserep.weizmann.ac.il} is also acknowledged. This research has made use of the NASA/IPAC Extragalactic Database (NED), which is operated by the Jet Propulsion Laboratory, California Institute of Technology, under contract with the National Aeronautics and Space Administration. This research made use of \textsc{RedPipe}\footnote{\url{https://github.com/sPaMFouR/RedPipe}} \citep{2021redpipe}, an assemblage of data reduction and analysis scripts written by AS.
\bibliography{17iro}
\bibliographystyle{aasjournal}
\end{document}